\documentclass[journal]{IEEEtran}

\IEEEoverridecommandlockouts
\usepackage{graphicx}
\usepackage{cite}
\usepackage{amsmath}
\usepackage{amsfonts}
\usepackage{amssymb}
\usepackage{subfigure}
\usepackage{psfrag}
\usepackage{color}
\usepackage{multicol}
\usepackage{tikz}
\usepackage{verbatimbox}
\usepackage{graphicx}
\usepackage{epsfig}
\usepackage{multirow}
\usepackage{lettrine}

\newcommand{\bm}{\mathbf}

\newcommand{\be}{\begin{equation}}
\newcommand{\ee}{\end{equation}}

\title{A Unified Framework for Pulse-Shaping on Delay-Doppler Plane}
\author{\normalsize Mohsen Bayat and Arman Farhang 
 \vspace{-0.45cm}
\thanks{Mohsen Bayat and Arman Farhang are with the Department of Electronic and Electrical Engineering, Trinity College Dublin, Dublin 2, Ireland
(Email: \{bayatm, arman.farhang\}@tcd.ie).
This publication has emanated from research conducted with the financial support of Science Foundation Ireland under Grant numbers SFI/19/FFP/7005(T) and SFI/21/US/3757. All calculations were performed on the Kelvin cluster maintained by the Trinity Centre for High Performance Computing. This cluster was funded through grants from the Higher Education Authority, through its PRTLI program.
}}

\begin{document}
\maketitle
\begin{abstract}
Delay-Doppler multiplexing has recently stirred a great deal of attention in research community. 
While multiple studies have investigated pulse-shaping aspects of this technology, it is challenging to identify the relationships between different pulse-shaping techniques and their properties. 
Hence, in this paper, we classify these techniques into two types, namely, circular and linear pulse-shaping.  
This paves the way towards the development of a unified framework that brings deep insights into the properties, similarities, and distinctions of different pulse-shaping techniques.
This framework reveals that the recently emerged waveform orthogonal delay-Doppler multiplexing (ODDM) is a linear pulse-shaping technique with an interesting staircase spectral behaviour. 
Using this framework, we derive a generalized input-output relationship that captures the influence of pulse-shaping on the effective channel. 
We also introduce a unified modem for delay-Doppler plane pulse-shaping that leads to the proposal of fast convolution based low-complexity structures. 
Based on our complexity analysis, the proposed modem structures are substantially simpler than the existing ones in the literature. Furthermore, we propose effective techniques that not only reduce the out-of-band (OOB) emissions of circularly pulse-shaped signals but also improve the bit-error-rate (BER) performance of both circular and linear pulse-shaping techniques.
Finally, we extensively compare different pulse-shaping techniques using various performance metrics.
\end{abstract}
\vspace{-0.45cm}
\section{introduction}\label{sec:Introduction}

\IEEEPARstart{I}{nitiated} by the landmark paper of Hadani et al. \cite{Hadani2017}, delay-Doppler multiplexing is a paradigm-shifting technology which has recently attracted a great deal of attention. 
The advantage of delay-Doppler multiplexing lies in exploiting the full channel diversity to gain resilience to the time-varying channel effects \cite{Hadani2018}. 
A common aspect of the emerging applications like autonomous vehicles, mmWave communications, and space-air-ground integrated networks is the highly dynamic wireless environment. While orthogonal frequency division multiplexing (OFDM) has been the technology of choice in the last two generations of wireless systems, it cannot cope with such time-varying wireless channels \cite{Wei2021}.

Orthogonal time-frequency space (OTFS) modulation is a foundational delay-Doppler multiplexing technique which can be implemented on the top of any time-frequency modulator \cite{Hadani2017}, e.g., OFDM, generalized frequency division multiplexing (GFDM) \cite{Tiwari2020}, filter bank multicarrier (FBMC) \cite{Pishva2022}, and universal filtered multi-carrier (UFMC) \cite{Andrea2023}. 
Recently, a new modulation technique called orthogonal delay-Doppler multiplexing (ODDM) was introduced in \cite{Lin2022_1, Lin2022_2, Lin2023_2}. 
ODDM can be viewed as a multicarrier system whose signal is formed by staggering multiple pulse-shaped OFDM symbols.
Since, backward compatibility is an important design aspect of future networks, our focus in this paper is on OFDM-based delay-Doppler multiplexing techniques, \cite{Fred2018, Raviteja2019b, Raviteja2018b, Zhou2023, Lin2022_0, Yao2021, Tong2023, Wei2021_2, Tusha2023, Farhang2018, Franz2022, Saifkhan2021, Saifkhan2022, Lin2022_1, Lin2022_2, Lin2023_2, Lin2022_3, Li2023}. 

While there is a large body of literature on this topic, \cite{Wei2023}, only a small number of studies focus on practical aspects such as pulse-shaping \cite{Wei2021_2, Tusha2023, Raviteja2019b, Raviteja2018b, Yao2021, Tong2023, Zhou2023, Lin2022_0}.
In \cite{Fred2018}, the authors show the superior performance of OTFS compared to OFDM with rectangular pulse-shapes.
Opposed to \cite{Raviteja2019b} that considers ideal pulse-shaping, the authors in \cite{Raviteja2018b} study the impact of rectangular pulse-shaping on the effective channel of OTFS. 
Subsequently, ideal and rectangular pulse-shaping techniques are compared in \cite{Zhou2023}.
The authors in \cite{Lin2022_0} reveal that rectangular pulses in OTFS lead to increased interference on the edges of the delay blocks at the receiver. 
The influence of the Nyquist transmit and receive filters on the effective channel for OTFS is investigated in \cite{Yao2021}. Similarly, the authors in \cite{Tong2023} derive an approximate effective channel for ODDM that resembles to the one in \cite{Yao2021}. It is worth noting that in both \cite{Yao2021} and \cite{Tong2023}, the transmit and receive filters are combined into a Nyquist one which is an approximation when the channel is time-varying. 

Beyond studying the pulse-shaping effects on OTFS, pulse-shape design is studied in \cite{Wei2021_2} and \cite{Tusha2023}. The main idea of \cite{Wei2021_2} is designing an optimal receiver window across the time dimension for interference mitigation purposes. To reduce the Doppler-induced leakage at the receiver, a global windowing method was introduced in \cite{Tusha2023}.
With the main focus on the pulse-shaping technique, the authors in \cite{Farhang2018, Saifkhan2021, Saifkhan2022, Lin2022_3, Li2023, Franz2022} demonstrate that the two-stage OTFS implementation in \cite{Hadani2017} can be effectively performed in one stage. This is made possible by using orthogonal delay-Doppler domain pulse-shapes based on the Zak transform theory \cite{Franz2022, Saifkhan2021}. 

Despite the existing literature on delay-Doppler multiplexing, there is a lack of a generalized framework that can represent different pulse-shaping techniques under the same umbrella. 
As mentioned earlier, the effective channel derivations that capture the influence of pulse-shaping involve approximations. Additionally, in the earlier literature on delay-Doppler multiplexing, oversampling has been disregarded.  
In reality, practical implementation of the transmit/receive pulse-shaping filters requires sampling above the Nyquist rate \cite{PS2007}. Hence, multi-stage digital oversampling is a necessity for practical implementation. From implementation viewpoint, the existing modem structures for delay-Doppler multiplexing are either complex or require approximations \cite{Lin2023_2}.

To address the aforementioned gaps in literature, in this paper, we classify the OFDM-based pulse-shaping techniques on the delay-Doppler plane into two categories of linear and circular pulse-shaping. 
We present the discrete-time formulation of these two classes with oversampling.  
Using this formulation, we derive a unified framework that sheds light on the properties, differences, and similarities of various pulse-shaping techniques. 
Based on this framework, we introduce a generalized modem structure for pulse-shaping on the delay-Doppler plane.
We also derive effective channel expressions by casting our generalized framework into matrix form. This leads to generalized and compact input-output relationships with no approximation.

As circular pulse-shaping does not allow the transients of the pulse to appear at the boundaries of the delay blocks, it has large out-of-band (OOB) emissions. 
Thus, we propose effective techniques that can substantially reduce the OOB emissions of circularly pulse-shaped signals. 
In particular, we propose OOB reduction techniques based on windowing and zero-guard (ZG) insertion both along the delay dimension to smooth the edges of the transmit signal. 
Based on our simulations, by investing a very small number of ZGs or samples to accommodate windowing, i.e., less than $5\%$ of the delay block length, around $20$~dB reduction in OOB emissions is achieved. As mentioned earlier, the channel imposed interference is severe at the edges of the delay blocks. 
Hence, inserting guard symbols at the edges of the delay blocks also leads to the bit-error-rate (BER) performance improvement for both circular and linear pulse-shaping techniques.
In this study, we also establish a discrete-time representation of the recently emerged waveform ODDM. 
Our derivations reveal that ODDM is a linear pulse-shaping technique. 
Furthermore, we present ODDM within the context of our unified framework which clearly explains the observed staircase behavior of its spectrum. 
To the best of our knowledge, such spectral behavior for ODDM has not been previously reported in the literature. 

As another contribution of this paper, using the above mentioned framework, we propose low-complexity modem structures for the pulse-shaping techniques under study. 
Using our proposed fast convolution based implementation of pulse-shaping, a substantial amount of complexity reduction is achieved. 
For instance, our proposed modem structure for ODDM leads to over two orders of magnitude complexity reduction compared to the proposed implementation in \cite{Lin2022_2}. Finally, we numerically confirm our mathematical derivations and claims through simulations.  
We thoroughly compare and analyze different puls-shaping techniques in terms of computational complexity, and performance metrics such as OOB emissions, BER and peak-to-average power ratio (PAPR).

The rest of this paper is organized as follows. Section~{\ref{sec:Principle}} presents the delay-Doppler plane multiplexing principles. Cyclic and linear pulse-shaping techniques are formulated in Sections~{\ref{sec:CPS}} and {\ref{sec:LPS}}, respectively, and the derivations therein are leveraged towards introduction of our unified pulse-shaping framework in Section~{\ref{sec:FW}}. 
Section~{\ref{sec:Input-Output}} presents the generalized input-output expressions for the pulse-shaping techniques under study.
Our proposed low-complexity modem structures are introduced in Section~\ref{sec:Comp}.
Our numerical results are discussed in Section~\ref{sec:Numerical} . Finally, Section~\ref{sec:Conclusion} concludes the paper.

\textit{Notations}:
Scalar values, vectors, and matrices are denoted by normal letters, boldface lowercase, and boldface uppercase, respectively.
$\mathbb{C}^{M \times N}$ denotes the set of $M \times N$ complex matrices.
$\bm{I}_{N}$ and $\bm{0}_{N}$ are $N \times N$ identity and zero matrices, respectively.
The notation $\bm{X}=\mathcal{T}_{M \times N}\{\bm{x}\}$ for an $(M+N-1) \times 1$ vector $\bm{x}$, represents an $M \times N$ Toeplitz matrix, in which $[\bm{X}]_{mn}=[\bm{x}]_{m-n+N}$. Similarly, $\bm{C}=\rm{circ}\{\bm{c}\}$ represents a circulant matrix whose first column is $\bm{c}$.
The superscripts $(.)^{\rm{H}}$, $(.)^{\rm{T}}$ and $(.)^{-1}$ indicate hermitian, transpose, and inverse operations, respectively.
$(\!(\cdot)\!)_N$ is the modulo-$N$ operator and $\mathcal{R \{ \cdot \}}$ denotes the real part of a complex number. 
$\otimes$, $*$ and $\stackrel{\mbox{\tiny{$M$}}}{\circledast}$ denote Kronecker product, linear convolution and $M$-point circular convolution, respectively.
$L$-fold upsampling/downsampling of a multidimensional signal along a given dimension, $l$, is represented by the operator $( \cdot )_{l,\uparrow_{L}}~/~( \cdot )_{l,\downarrow_{L}}$.
$N$-point discrete Fourier transform (DFT) of the 2D signal $x[l,n]$ with respect to $n$, is denoted as $X[l,k]=\mathcal{F}_{N,n} \{ x[l,n] \} = \frac{1}{\sqrt{N}}\sum_{n=0}^{N-1} x[n] e^{-j\frac{2\pi kn}{N}}$.
Finally, $\delta[\cdot]$ is the Dirac delta function.

 \vspace{-0.2cm}
\section{Delay-Doppler Plane Multiplexing Principles}\label{sec:Principle}

We consider the quadrature amplitude modulated (QAM) data symbols $D[l,k]$ on a regular grid in the delay-Doppler domain with $M$ delay bins and $N$ Doppler bins, where $l=0,\ldots, M-1$ and $k=0,\ldots, N-1$. Given the delay spacing of $\Delta \tau$, each delay block with $M$ delay bins has the duration of $T=M\Delta \tau$. Hence, the Doppler spacing is $\Delta \nu = \frac{1}{NT}$.
To transmit the data symbols over the wireless channel, as it was proposed in \cite{Hadani2017}, they are first translated into the time-frequency domain using the inverse symplectic finite Fourier transform (ISFFT) operation. Then, they are fed into a time-frequency modulator, such as OFDM.
ISFFT involves application of $M$-point discrete Fourier transform (DFT) and $N$-point inverse DFT (IDFT) operations across the delay and Doppler dimensions, respectively, i.e.,
\begin{align} \label{eqn:isfft} \mathcal{X}[m,n] = \frac{1}{\sqrt{MN}} \sum_{l=0}^{M-1}\sum_{k=0}^{N-1} D[l,k] e^{j2 \pi (\frac{kn}{N}-\frac{lm}{M})}, \end{align}
where $m=0,\ldots,M-1$ and and $n=0,\ldots,N-1$ are the frequency and time indices, respectively.
The resulting signal $\mathcal{X}[m,n]$ is then passed through an OFDM modulator to obtain the delay-time domain signal 
\be\label{eqn:X_ln}
X[l,n] = \frac{1}{\sqrt{M}} \sum_{m=0}^{M-1} \mathcal{X}[m,n] e^{j \frac{2 \pi ml}{M}}.
\ee
To form the transmit signal, the samples on the delay-time grid, $X[l,n]$, are first converted to a serial stream $x[\kappa]$, where $x[\kappa]=X[l,n]$ for $\kappa=nM+l$, $n=0,\ldots,N-1$, and $l=0,\ldots,M-1$. 
Then, a cyclic prefix (CP) with length $L_{\rm{cp}} \geq L_{\rm{ch}}$ is appended at the beginning of $x[\kappa]$, where $L_{\rm{ch}}$ is the channel delay spread. The CP is inserted to avoid inter-block interference.
Therefore, the transmit signal is represented as
\be\label{eqn:s[k]}
s[\kappa] = 
\begin{cases}
x[(\!(\kappa)\!)_{MN}], &\text{$-L_{\rm{cp}}\leq\kappa\leq MN-1$}, \\
0, &\text{otherwise.}
\end{cases}
\ee

After $s[\kappa]$ is transmitted through the linear time-varying (LTV) channel, the received signal can be expressed as
\begin{align} \label{eqn:rec} r[\kappa] = \sum_{i=0}^{L_{\rm{ch}}-1} h[\kappa,i] s[\kappa-i] + \eta[\kappa], \end{align}
where $h[\kappa,i]$ is the channel impulse response at the delay tap $i$ and the sample $\kappa$, and $\eta[\kappa] \sim \mathcal{CN} (0,\sigma^2_{\eta})$ is the complex additive white Gaussian noise (AWGN) with the variance $\sigma^2_{\eta}$. To obtain the received symbols in the delay-Doppler domain, the CP is first removed from $r[\kappa]$. Considering the remaining $MN$ samples of $r[\kappa]$, i.e., $\kappa=0,\ldots,MN-1$, on a regular 2D grid, the received delay-time domain signal is formed as $Y[l,n]=r[nM+l]$ with the delay and time indices $l=0,\ldots, M-1$ and $n=0,\ldots, N-1$, respectively.
Afterwards, $Y[l,n]$ is passed into an OFDM demodulator to find the time-frequency domain signal
\be\label{eqn:ymn} \mathcal{Y}[m,n] = \frac{1}{\sqrt{M}}\sum_{l=0}^{M-1} Y[l,n] e^{-j \frac{2 \pi ml}{M}}. \ee
 The received delay-Doppler domain data symbols are then obtained by performing an SFFT operation on $\mathcal{Y}[m,n]$, i.e.,
\be\label{eqn:Dlk}
\widetilde{D}[l,k] = \frac{1}{\sqrt{MN}} \sum_{m=0}^{M-1}\sum_{n=0}^{N-1} \mathcal{Y}[m,n] e^{-j2 \pi (\frac{kn}{N}-\frac{lm}{M})}.
\ee
Finally, the transmit data symbols are estimated by passing the received signal, $\widetilde{D}[l,k]$, through a channel equalizer.  

As with any communication system, pulse-shaping is an important design consideration for delay-Doppler multiplexing that requires careful attention. Therefore, in the rest of the paper, we categorize, analytically study and compare different pulse-shaping techniques for delay-Doppler multiplexing. In particular, we derive a unified framework that represents different pulse-shaping options under the same umbrella, reveals their spectral properties and illustrates the underlying relationship between them.

\section{Cyclic Pulse-Shaping on Delay-Doppler Plane}\label{sec:CPS}

The focus of this section is on oversampling and pulse-shaping the multiplexed data in delay-Doppler domain by using the OFDM modulator. For reasons that will become clear shortly, this type of pulse-shaping can be categorized as a circular pulse-shaping technique. However, OFDM-based pulse-shaping leads to a high level of OOB emissions. Therefore, in Sections~\ref{subsec:OOB reduction} and \ref{subsec:OOB reduction ZP}, we propose effective OOB emission reduction techniques to tackle this issue.

\vspace{-0.3cm}
\subsection{Cyclic/Circular Pulse-shaping}\label{subsec:CPS-OTFS}
As discussed in Section~\ref{sec:Principle}, after converting the delay-Doppler domain data symbols to the time-frequency domain, they can be transmitted using the OFDM modulator. Thus, oversampling and pulse-shaping can be performed by the OFDM modulator. Given the oversampling factor $L_{\rm{us}}$, this process is carried out by increasing the number of frequency bins from $M$ to $M'\!=\!M\!L_{\rm{us}}$ in each time slot \cite{Farhang2010}. Considering the ideal interpolation filter, the additional $M'-M$ high-frequency bins are set to zero and then an $M'$-point IDFT is applied to the resulting signal. 

\begin{figure*}[!t]
  \centering 
  {\includegraphics[scale=0.088]{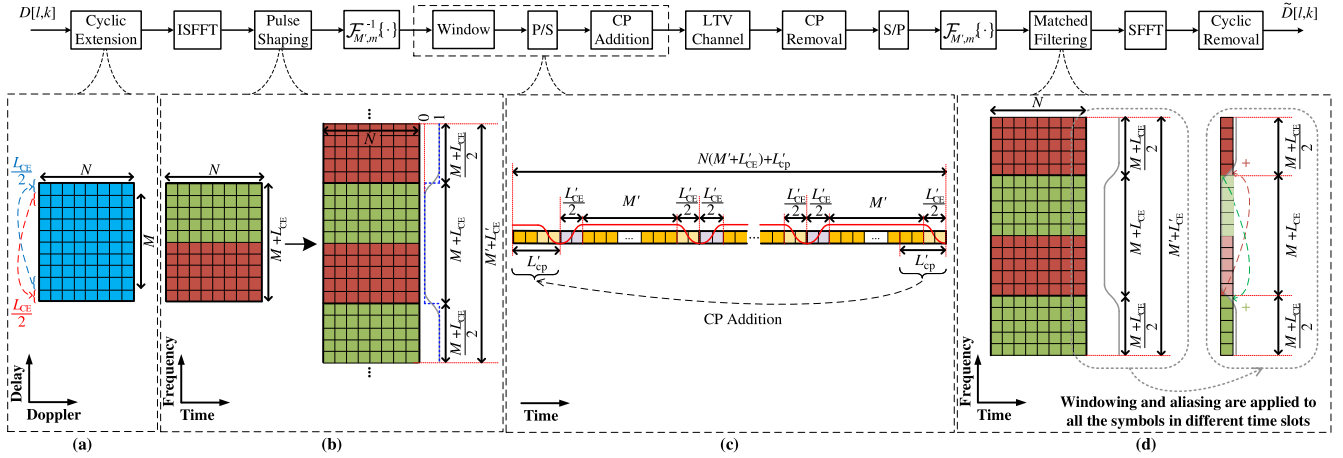}}
  \vspace{-0.2cm}
  \caption{Modem structure for C-PS OTFS with CE before pulse-shaping when $M=8$, $N=8$, $L_{\rm{us}}=2$, $L_{\rm{CE}}=2$.} \vspace{-0.4cm}
  \label{fig:wola}
\end{figure*}

In a more general form, frequency domain interpolation can be implemented in two steps; (i) generation of $L_{\rm{us}}$ spectral replicas of $\mathcal{X}[m,n]$ at each time slot $n$ for $m=0,\ldots,M-1$ to form the periodic sequence $\mathcal{X}_{\rm p}[m',n]=\mathcal{X}[(\!(m')\!)_{M},n]$ for $m'=0,\ldots,M'-1$, and (ii) multiplication of the resulting sequence in each time slot, $n$, by the frequency response of an interpolation filter, see Fig.~\ref{fig:wola}~(b).
Here, we deploy the square-root raised-cosine (SRRC) interpolation filter, with the frequency response
\be\label{eqn:rrc_freq}
\psi_{\alpha}[m'] \!=\!\!
\begin{cases}\!
\frac{1}{\sqrt{M}} f_{\alpha}[m'], & \!\!\text{ $0 \!\leq\! m' \!\! \leq \!\! \alpha M\!$ \& $\!M\! \leq \!\! m' \!\! \leq \!\! (1\!+\!\alpha)M$,} \\
\frac{1}{\sqrt{M}}, & \!\!\text{$\alpha M \leq m' \leq M$,} \\
0, & \!\!\text{otherwise,}
\end{cases}
\ee
where $f_{\alpha}[m'] \!=\! \frac{1}{\sqrt{2}} [1 + \cos(\frac{\pi}{\alpha} (\frac{|m'-\lfloor \frac{(1+\alpha)M}{2} \rfloor|}{M} - \frac{1-\alpha}{2}))]^{\frac{1}{2}}$ and $\alpha$ is the roll-off factor.
The amount of excess bandwidth and equivalently the sidelobes of the interpolation filter in delay domain are controlled by the roll-off factor. For $\alpha=0$, $\psi_0[m']$ is the ideal brick-wall filter. 

After frequency domain interpolation, the oversampled signal in the delay-time domain, i.e., the signal at the output of the OFDM modulator, is obtained as
\be \label{eqn:X_bar} \overline{X}^{\mathtt{C}}[l',n] = \frac{1}{\sqrt{M'}} \sum_{m'=0}^{M'-1} \overline{\psi}_{\alpha}[m']\mathcal{X}_{\rm p}[m',n] e^{j \frac{2 \pi m'l'}{M'}}, \ee
where $\overline{\psi}_{\alpha}[m']\!=\!\psi_{\alpha}[(\!(m'\!+\!\lfloor \!\frac{(1+\alpha)M}{2}\! \rfloor)\!)_{M'}]$,
for $l'\!=\!0,\ldots,M'-1$ and $n=0,\ldots,N-1$. This signal is considered as the pulse-shaped OTFS signal. 

The above oversampling process can be equivalently performed in the delay-time domain in the following two steps; (i) $L_{\rm{us}}$-fold expansion of the delay-time domain signal $X[l,n]$, in (\ref{eqn:X_ln}), for each time slot $n$, to produce the sequence 
\be \label{eqn:us} 
X_{\rm{e}}[l',n] = \sum_{l=0}^{M-1}X[l,n] \delta[l'-lL_{\rm{us}}],
\ee
and (ii) the $M'$-point circular convolution of the resulting signal, along the delay dimension, with the impulse response of the interpolation filter to obtain
\begin{align} \label{eqn:cotfs_ps} \overline{X}^{\mathtt{C}}[l',n] = X_{\rm{e}}[l',n] \stackrel{\mbox{\tiny $~~M'$}}{\circledast} p[l'], \end{align} 
where $p[l'] = \mathcal{F}^{-1}_{M',m'}\{\overline{\psi}_{\alpha}[m']\} = \frac{{\rm{sinc}}(l') \cos(\pi \alpha l')}{1- (2 \alpha l')^2}$ is a square-root Nyquist filter which can be considered as the pulse-shaping filter. The superscript `$\mathtt{C}$' in (\ref{eqn:cotfs_ps}) is used to emphasize the type of pulse-shaping, i.e., circular.

The transmit signal before CP addition can be formed by turning the two-dimensional delay-time domain signal, $\overline{X}^{\mathtt{C}}[l',n]$, into the serial stream $x_{\rm{us}}^{\mathtt{C}}[\kappa']=\overline{X}^{\mathtt{C}}[l',n]$ where $\kappa'=nM'+l'$.
Considering $\overline{X}^{\mathtt{C}}[l',n]$ to be non-zero only for $l'=0,\ldots, M'-1$ and $n=0,\ldots, N-1$ and zero otherwise, $x_{\rm{us}}^{\mathtt{C}}[\kappa']$ can be represented as
\begin{align} \label{eqn:cotfs_p/s} x_{\rm{us}}^{\mathtt{C}}[\kappa'] = \sum_{n=0}^{N-1} \overline{X}^{\mathtt{C}}[\kappa'-nM',n]. \end{align}
By substituting (\ref{eqn:cotfs_ps}) into (\ref{eqn:cotfs_p/s}), we have
\begin{align} \label{eqn:cotfs_p/s2} x_{\rm{us}}^{\mathtt{C}}[\kappa'] &= \sum_{n=0}^{N-1} X_{\rm{e}}[\kappa'-nM',n] \stackrel{\mbox{\tiny{$~~M'$}}}{\circledast} p[\kappa'-nM']. \end{align}
Substituting (\ref{eqn:X_ln}) into (\ref{eqn:us}) and changing the summation order, one may realize that $X_{\rm{e}}[l',n]$ can be directly obtained from expanding the delay-Doppler domain data symbols, i.e.,
\begin{align} \label{eqn:cotfs_p/s3}
X_{\rm{e}}[l',n] &\!=\! \sum_{l=0}^{M\!-\!1} \! \left( \frac{1}{\sqrt{N}} \sum_{k=0}^{N\!-\!1} D[l,k] e^{j \frac{2 \pi nk}{N}} \right) \delta[l'-lL_{\rm{us}}] \nonumber \\ &= \frac{1}{\sqrt{N}} \sum_{k=0}^{N-1} \left( \sum_{l=0}^{M-1} D[l,k] \delta[l'-lL_{\rm{us}}] \right) e^{j \frac{2 \pi nk}{N}} \nonumber \\ &= \!\frac{1}{\sqrt{N}}\!  \sum_{k=0}^{N-1}\! D_{\rm{e}}[l',k] e^{j \frac{2 \pi nk}{N}} \!\!=\! \mathcal{F}_{N,k}^{-1} \Bigl\{ D_{\rm{e}}[l',k] \Bigr\}, \end{align}
where $D_{\rm{e}}[l',k]=\Big( D[l,k] \Big)_{l,\uparrow_{L_{\rm us}}}$.
Using (\ref{eqn:cotfs_p/s3}) in (\ref{eqn:cotfs_p/s2}), we have
\begin{align} \label{eqn:cotfs_p/s4} x_{\rm{us}}^{\mathtt{C}}[\kappa'] &= \!\! \sum_{n=0}^{N-1} \mathcal{F}_{N,k}^{-1} \Bigl\{ D_{\rm{e}}[\kappa'\!-\!nM',k] \Bigr\} \stackrel{\mbox{\tiny{$~~M'$}}}{\circledast} p[\kappa'\!-\!nM'] \nonumber\\ &= \!\! \sum_{n=0}^{N-1} \! \mathcal{F}_{N,k}^{-1} \Bigl\{ D_{\rm{e}}[\kappa'\!-\!nM',k] \stackrel{\mbox{\tiny{$~~M'$}}}{\circledast} p[\kappa'\!-\!nM'] \Bigr\}. \end{align}
Hence, the transmit signal can be oversampled via expanding and circular filtering operations along the delay dimension in either the delay-Doppler domain or the delay-time domain. The circular convolution in (\ref{eqn:cotfs_p/s4}) can also be interpreted as making $D_{\rm{e}}[l',k]$ and $p[l']$ periodic along the delay dimension, linearly convolving them, and then confining the resulting signal with a length $M'$ rectangular pulse. This can be thought of as a circular pulse-shaping procedure for OTFS which we call C-PS OTFS. After appending a CP with length $L'_{\rm{cp}}\!=\!L_{\rm{us}}L_{\rm{cp}}$ to the beginning of the OTFS signal, it is transmitted through the channel.

Considering the oversampling factor $L_{\rm{us}}$ at the receiver, the received signal after CP removal is represented as $r_{\rm{us}}^{\mathtt{C}}[\kappa']$. Then, we form the 2D delay-time domain signal as $Y_{\rm{us}}^{\mathtt{C}}[l',n]=r_{\rm{us}}^{\mathtt{C}}[l'+nM']$ where $l'=0,\ldots,M'-1$ and $n=0,\ldots,N-1$.
Since $p[l']$ is a square-root Nyquist filter, in an ideal channel, the transmit data symbols can be perfectly reconstructed by matched filtering and downsampling the received signal along the delay dimension. Consequently, the received delay-Doppler domain data symbols, which are affected by the channel, can be obtained by performing the transmitter operations in reverse order via two steps: (i) filtering the received signal with $p^*[-l']$ along the delay dimension, i.e., matched filtering, and (ii) downsampling the matched filter output by the factor $L_{\rm{ds}}=L_{\rm{us}}$, followed by $N$-point DFT operation along the time dimension. Similar to the transmitter, the operations along the delay dimension, at the receiver, can be performed in either the delay-Doppler or delay-time domain, i.e.,
\begin{align} \label{eqn:D_tilde} \widetilde{D}[l,k] &= \mathcal{F}_{N,n} \Bigl\{ \Big(Y_{\rm{us}}^{\mathtt{C}}[l',n] \stackrel{\mbox{\tiny{$~~M'$}}}{\circledast} p[l']\Big)_{l',\downarrow_{L_{\rm{ds}}}}\Bigr\} \nonumber\\ &=  \Big(\mathcal{F}_{N,n} \Bigl\{  Y_{\rm{us}}^{\mathtt{C}}[l',n] \Bigr\} \stackrel{\mbox{\tiny{$~~M'$}}}{\circledast} p[l']\Big)_{l',\downarrow_{L_{\rm{ds}}}}, \end{align}
where $p^*[-l']=p[l']$ as $p[l']$ is real-valued and symmetric.

The above process at the receiver can alternatively be performed in the time-frequency domain, where circular convolution reduces to a simple multiplication operation across frequency. 
To this end, $Y_{\rm{us}}^{\mathtt{C}}[l',n]$ is first translated into the time-frequency domain by taking an $M$-point DFT along the delay dimension, i.e., ${\mathcal{Y}}_{\rm{us}}[m',n]=\frac{1}{\sqrt{M'}} \sum_{l'=0}^{M'-1} Y_{\rm{us}}^{\mathtt{C}}[l',n] e^{-j \frac{2 \pi l'm'}{M'}}$. Matched filtering and downsampling is then carried out in the frequency domain in two steps; (i) multiplying the frequency response of the decimation filter, $\overline{\psi}_{\alpha}[m']$ for $\forall n$, to ${\mathcal{Y}}_{\rm{us}}[m',n]$, and (ii) aliasing the resulting signal in the frequency dimension at each time slot to obtain ${\mathcal{Y}}[m,n]$, see Fig.~\ref{fig:wola}~(d). Based on the sampling theory principles, \cite{proakis2007digital}, aliasing in the frequency domain is equivalent to downsampling in the time domain.
The received delay-Doppler domain symbols are then obtained using (\ref{eqn:Dlk}).

Despite simple implementation of C-PS OTFS in the time-frequency domain, circular pulse-shaping has large OOB emissions. This is due to the fact that circular convolution is confined to a rectangular window. Hence, circular pulse-shaping does not allow the transients of the pulse-shape to appear which leads to abrupt changes at the boundaries of the delay blocks.
To tackle this issue, in the following subsections, we introduce effective techniques that can substantially reduce the OOB emissions of C-PS OTFS.

\vspace{-0.3cm}
\subsection{Proposed OOB reduction technique by windowing}\label{subsec:OOB reduction}
A classical approach in reducing the OOB emissions is to make the edges of each time slot (delay block) smooth using windowing techniques \cite{AFarhang2010}. Consequently, one may choose to feed the time-frequency signal, $\psi_{\alpha}[m']\mathcal{X}_{\rm p}[m',n]$, into a windowed OFDM transmitter. To facilitate the windowing operation, the output of the OFDM modulator, i.e., the delay-time domain signal $\overline{X}^{\mathtt{C}}[l',n]$ from (\ref{eqn:X_bar}), is first cyclically extended at both sides along the delay dimension. Then, the window is applied to the resulting signal, i.e., $\overset{\huge\frown}{X}[\ell',n]=\overline{X}^{\mathtt{C}}[(\!(\ell')\!)_{M'},n]$ where $\ell'=-\frac{L'_{\rm{CE}}}{2}, \ldots, M'+\frac{L'_{\rm{CE}}}{2}-1$ and $L'_{\rm{CE}}$ is the cyclic extension (CE) length. As a representative example, for a raised-cosine (RC) window with the roll-off factor $\beta$, a CE of length $L'_{\rm{CE}}\!=\!\lfloor {\beta M}\rfloor L_{\rm{us}}$ samples is required. This procedure is widely deployed for OOB emission reduction in OFDM transmitters \cite{Farhang2010}. As it is shown in Fig.~\ref{fig:wola}~(c), a CP is appended to the beginning of the block after the windowing operation.

After CP removal and discarding the CEs at the receiver, the resulting signal is fed into the OFDM demodulator. The SFFT operation is then applied at the OFDM demodulator output and the received data symbols in the delay-Doppler domain are obtained. However, discarding the CEs before the signal is passed into the OFDM demodulator becomes problematic as the channel becomes increasingly time-varying. The authors in \cite{Zegrar2023} have briefly pointed out that the cyclic extensions within OTFS block contain meaningful data and should not be simply removed. However, not much detail on the effect of discarding the CEs in OTFS is provided in \cite{Zegrar2023}. Thus, in the following, we explain the underlying reasons behind the adverse effect of CE removal before feeding the signal into OFDM demodulator. This leads to our proposal of CE addition before oversampling the signal in the delay-Doppler domain. 

From sampling theory, \cite{proakis2007digital}, one may realize that CE addition at the transmitter along the delay dimension is equivalent to upsampling in the frequency domain. Conversely, CE removal at the receiver corresponds to rate reduction in the frequency domain. Therefore, when the channel does not have any spreading effects in the frequency domain and it is only frequency selective, i.e., time-invariant, no information is lost after discarding the CEs. In contrast, when the channel is time-varying, Doppler spread leads to frequency domain signal leakage. Thus, depending on the amount of Doppler spread, a certain amount of signal power is lost after CE removal which corresponds to downsampling in the frequency domain.

To avoid the aforementioned issues, we propose to increase the delay-Doppler grid size along the delay dimension to accommodate the CEs before oversampling and pulse-shaping. In particular, the grid size of $M\times N$ becomes $M_{\rm{CE}}\times N$ where $M_{\rm{CE}}=M+L_{\rm CE}$ and $L_{\rm{CE}}$ is the CE length. Consequently, we cyclically extend the data, $D[l,k]$, along the delay dimension to obtain $\overset{\huge\frown}{D}[\ell,k]=D[(\!(\ell)\!)_{M},k]$ where $\ell=-\frac{L_{\rm{CE}}}{2}, \ldots, M+\frac{L_{\rm{CE}}}{2}-1$. For the example of the RC window, the CE length is given by $L_{\rm{CE}}= \lfloor {\beta M} \rfloor$. 
Replacing $D[l,k]$ by $\overset{\huge\frown}{D}[\ell,k]$ in (\ref{eqn:isfft}) and using (\ref{eqn:X_bar}), 
the oversampled transmit signal in the delay-time domain, $\overline{X}^{\mathtt{C}}[\ell',n]$, is generated where $\ell'=-\frac{L'_{\rm{CE}}}{2}, \ldots, M'+\frac{L'_{\rm{CE}}}{2}-1$ and $L'_{\rm{CE}}=L_{\rm{CE}}L_{\rm{us}}$. After windowing along the delay dimension, we obtain
\be \label{eqn:wola_sort} x_{\rm{w}}^{\mathtt{C}}[\kappa'] \!=\!\!\! \sum_{n=0}^{N-1} \! \overline{X}^{\mathtt{C}}[\kappa'-\frac{L'_{\rm{CE}}}{2}-nM'_{\rm{CE}},n] \psi_{\beta,\rm{w}}[\kappa'-\frac{L'_{\rm{CE}}}{2}-nM'_{\rm{CE}}], \ee
where $\kappa' \!=\! 0,\ldots,NM'_{\rm{CE}}-1$, $n=0,\ldots,N-1$, $M'_{\rm{CE}}=M'+L'_{\rm{CE}}$, and
\be\label{eqn:rrc_time}
\psi_{\beta,\rm{w}}[\ell'] \!=\!\!
\begin{cases}
g_{\beta}[\ell'], & \text{$0 \leq \ell' \leq \frac{\beta}{2} M'$ \!\&\! $M' \leq \ell' \leq (1\!+\!\frac{\beta}{2})M'$,} \\
1, & \text{$\frac{\beta}{2} M' \leq \ell' \leq M'$,} \\ 
0, & \text{otherwise,}
\end{cases}
\ee
given $g_{\beta}[\ell']\!=\!\frac{1}{2}[1+\cos(\frac{2 \pi}{\beta}(\frac{|\ell'-\lfloor \frac{(1+\frac{\beta}{2})M'}{2} \rfloor|}{M'}-\frac{1-\frac{\beta}{2}}{2}))]$. 
Then, a CP with length $L'_{\rm cp} = L_{\rm cp}L_{\rm{us}}$ is appended at the beginning of $x^{\mathtt{C}}_{\rm{w}}[\kappa']$ and the resulting signal is transmitted through the channel. 

At the receiver side, we follow the same procedure as in Section~\ref{subsec:CPS-OTFS} to obtain the received signal in delay-Doppler domain where we can safely remove the CEs without incurring any performance penalty.

\subsection{Proposed OOB reduction technique by zero-guards (ZGs)}\label{subsec:OOB reduction ZP}
As mentioned earlier, the cyclic convolution in (\ref{eqn:cotfs_ps}) does not allow the smooth transients of the pulse-shape to appear at the boundaries of each time-slot. This is because the transients wrap-around at the edges of each delay block. To avoid the wrap-around effect, we propose zero-padding the delay-Doppler domain data symbols in all time-slots, along the delay dimension. As an alternative to the windowing technique, zero-padding leads to the appearance of smooth transients, at the edges of each delay block. Consequently, based on our simulation results in Section~\ref{sec:Numerical}, a substantial reduction in OOB emissions is achieved by inserting only a small number of zero delay bins, $L_{\rm{ZG}}$, as guards. Furthermore, based on the results in \cite{Raviteja2018b} and \cite{Lin2022_0}, the interference between the adjacent delay blocks due to the channel delay spread is more severe in the delay bins at the edges of each block. Hence, ZGs not only improve the spectral containment of the transmit signal but they also lead to an improved BER performance which is shown by simulations in Section~\ref{sec:Numerical}.

\section{Linear Pulse-Shaping on Delay-Doppler Plane}\label{sec:LPS}

Based on the discussions in Section~\ref{sec:CPS}, oversampling and pulse-shaping the translated data symbols to the time-frequency domain using OFDM modulator, is equivalent to cyclic pulse-shaping. However, this type of pulse-shaping leads to large OOB emissions which can be alleviated through windowing or inserting ZGs along the delay dimension, as compromise choices. 
Even though a small number of samples are required as the CEs for windowing or as the guard symbols, they lead to bandwidth efficiency loss. Hence, in this section, we explain how this loss can be avoided by linear pulse-shaping while achieving low OOB emissions. Furthermore, we present the discrete-time formulation of the recently emerged delay-Doppler domain multiplexing technique, ODDM. This formulation shows that ODDM is a linear pulse-shaping technique while revealing its interesting properties that have not been reported in the literature thus far.

\subsection{Linear Pulse-shaping} \label{subsec:LPS-OTFS}
As explained in Section~\ref{subsec:CPS-OTFS}, frequency domain oversampling is performed in two stages. At the first stage, the translated data symbols to the time-frequency domain are repeated $L_{\rm us}$ times across frequency. At the second stage, only one replica is preserved by cyclic filtering, see (\ref{eqn:cotfs_ps}). Alternatively, in this section, we remove the unnecessary spectral replicas by linear filtering. Linear filtering operation makes the signal edges in each time-slot smooth which leads to a significant reduction in OOB emissions. However, the filter transients at the edges of each time-slot lead to spectral efficiency loss. Since, $p[l']$ is a square-root Nyquist filter, matched filtering at the receiver leads to a Nyquist pulse with zero-crossings at the delay spacings. Therefore, to avoid bandwidth efficiency loss, similar to FBMC systems, \cite{Farhang2010}, the adjacent time-slots are allowed to overlap.

Accordingly, linear pulse-shaping for a given time-slot, $n$, can be performed by replacing circular convolution with linear convolution in equation (\ref{eqn:cotfs_ps}), i.e.,
\begin{align} \label{eqn:lotfs_ps} \overline{X}^{\mathtt{L}}[l',n] = X_{\rm{e}}[l',n] * p[l']. \end{align}
where $l'=-\frac{M'}{2}+1,\ldots, M'+\frac{M'}{2}-1$. Consequently, the transmit signal before appending the CP can be formed by overlapping the adjacent time-slots with the spacing of $M'$ samples as
\begin{equation} \label{eqn:lotfs_p/s} x^{\mathtt{L}}_{\rm{us}}[\kappa'] = \sum_{n=0}^{N-1} \overline{X}^{\mathtt{L}}[\kappa'-nM',n], \end{equation}
where $\kappa'\!=\!-\frac{M'}{2}+1,\ldots, M'N\!+\!\frac{M'}{2}\!-\!1$. It is worth noting that the first ${M'}/{2}-1$ and the last ${M'}/{2}$ samples of $x_{\rm{us}}^{\mathtt{L}}[\kappa']$ are due to the transient interval of the pulse-shape $p[l']$. These intervals are similar to ramp-up and ramp-down segments in FBMC signals \cite{FarhangB2016}. This can result in significant bandwidth efficiency loss when the number of time-slots, $N$, is small. To reduce the ramp-up and ramp-down duration, the pulse $p[l']$ can be truncated up to $Q$ zero-crossings that include significant sidelobes at each side of its main lobe, \cite{Lin2022_2}. This comes at the expense of an approximation in satisfying the Nyquist criterion. Hence, perfect reconstruction of the transmit symbols at the receiver cannot be achieved, \cite{Farhang2010}. While pulse truncation does not affect the BER performance for small constellation sizes, as the constellation size increases, it may lead to performance penalty. Furthermore, for very small choices of $Q$, truncation results in significantly increased OOB emissions. Consequently, $Q$ should be carefully chosen to avoid or at least alleviate these unfavourable effects.

Using (\ref{eqn:cotfs_p/s3}) and (\ref{eqn:lotfs_ps}), equation (\ref{eqn:lotfs_p/s}) can be rearranged as 
\begin{align} \label{eqn:lotfs_p/s3} x_{\rm{us}}^{\mathtt{L}}[\kappa'] &=\sum_{n=0}^{N-1} X_{\rm{e}}[\kappa'-nM',n] * p[\kappa'-nM'] \nonumber\\ &= \sum_{n=0}^{N-1} \mathcal{F}_{N,k}^{-1} \Bigl\{ D_{\rm{e}}[\kappa'\!-\!nM',k] \Bigr\} * p[\kappa'\!-\!nM'] \nonumber\\ &= \sum_{n=0}^{N-1} \! \mathcal{F}_{N,k}^{-1} \Bigl\{ D_{\rm{e}}[\kappa'\!-\!nM',k] * p[\kappa'\!-\!nM'] \Bigr\}, \end{align}
where $p[l']$ is the truncated pulse with the parameter $Q$, $\kappa'=-Q',\ldots, M'N+Q'-1$, and $Q'=QL_{\rm{us}}$ for $Q\in\{1,\ldots,\frac{M}{2}\}$. From (\ref{eqn:lotfs_p/s3}), one may realize that oversampling and pulse-shaping can be performed either directly in the delay-Doppler domain or after the signal is converted to the delay-time domain. This is due to the fact that pulse-shaping is carried out across the delay dimension and independent of Doppler. After pulse-shaping, a CP with length $L'_{\rm cp}$ is placed at the beginning of the transmit signal. Since the transmit signal is formed based on the OTFS framework, we refer to this system as linearly pulse-shaped OTFS (L-PS OTFS).

Considering the oversampling factor $L_{\rm{us}}$ at the receiver, the received signal after CP removal can be represented as $r_{\rm{us}}^{\mathtt{L}}[\kappa']$. Thus, we rearrange the received signal samples to form the 2D signal $Y_{\rm{us}}^{\mathtt{L}}[l',n]=r_{\rm{us}}^{\mathtt{L}}[l'+nM']$ for $l'=-Q',\ldots,M'+Q'-1$ and $n=0,\ldots,N-1$ in delay-time domain. Using the reverse operations to those at the transmitter, the received delay-Doppler domain data symbols, affected by the channel, can be obtained in two steps. First, by matched filtering and then by downsampling with the factor $L_{\rm ds}=L_{\rm{us}}$ both along the delay dimension. This operation can be performed either in delay-time domain or after converting the signal $Y_{\rm{us}}^{\mathtt{L}}[l',n]$ to the delay-Doppler domain, i.e., 
\begin{align} \label{eqn:Dl_tilde} \widetilde{D}[l,k] &= \mathcal{F}_{N,n} \Bigl\{ \Big(Y_{\rm{us}}^{\mathtt{L}}[l',n] * p[l']\Big)_{l',\downarrow_{L_{\rm ds}}}\Bigr\} \nonumber\\ &=  \Big(\mathcal{F}_{N,n} \Bigl\{  Y_{\rm{us}}^{\mathtt{L}}[l',n] \Bigr\} * p[l']\Big)_{l',\downarrow_{L_{\rm{ds}}}}, \end{align}
recalling that $p^*[-l']=p[l']$.

\vspace{-0.2cm}
\subsection{Discrete-Time Presentation of ODDM}\label{subsection:oddm}
ODDM, \cite{Lin2022_1, Lin2022_2}, is a new delay-Doppler domain multiplexing technique that has recently emerged. This waveform directly utilizes an orthogonal delay-Doppler domain pulse for data transmission. To form the transmit signal, each data symbol, $D[l,k]$, scales the pulse-shape that is shifted by $l$ positions along the delay dimension and modulated to the Doppler frequency $k/NT$, i.e.,
\be \label{eqn:Cont_DT} x^{\mathtt{ODDM}}(t) = \sum_{l=0}^{M-1} \sum_{k=0}^{N-1} D[l,k] u(t-l\frac{T}{M}) e^{j \frac{2 \pi k}{NT}(t-l\frac{T}{M})}, \ee
where $-\frac{T}{2} \!\! \leq \!\! t \!\! < \!\! NT+\frac{T}{2}$, $T\!\!=\!\!M \Delta \tau$ is the symbol period, $u(t)=\sum_{n=0}^{N-1} p(t-nT)$ is the delay-Doppler domain pulse-shape, and $p(t)$ is a Nyquist pulse that can be truncated as mentioned earlier. After appending a CP to the beginning of $x^{\mathtt{ODDM}}(t)$, the resulting signal is transmitted over the channel.

At the receiver side, the CP is first discarded. Then, the transmit data symbols that are affected by the channel are obtained by matched filtering, \cite{Lin2022_2}. As ODDM has just emerged, it is still in its infancy. Therefore, in the following, we derive the discrete-time formulation of this waveform that sheds light on some of its unexplored aspects.

Considering the oversampling factor of $L_{\rm{us}}$, (\ref{eqn:Cont_DT}) can be represented in discrete time as 
\begin{align} \label{eqn:Disc_DT1}
x_{\rm{us}}^{\mathtt{ODDM}}[\kappa'] = &\sum_{l=0}^{M-1} \sum_{k=0}^{N-1} D[l,k] u[\kappa'-lL_{\rm{us}}] e^{j \frac{2\pi k(\kappa'-lL_{\rm{us}})}{NM'}},
\end{align}
where $\kappa'\!=\!-Q',\ldots,M'N+Q'-1$, $x_{\rm{us}}^{\mathtt{ODDM}}[\kappa'] \!=\! x^{\mathtt{ODDM}}(\kappa' \!\frac{T}{M'})$, $u[\kappa'-lL_{\rm{us}}]\!=\!u((\kappa'-lL_{\rm{us}})\frac{T}{M'})$, and $u[\kappa']\!=\!\!\sum_{n=0}^{N-\!1}\!p[\kappa'-nM']$.
Therefore, (\ref{eqn:Disc_DT1}) can be expanded as
\begin{align} \label{eqn:Disc_DT2}
x_{\rm{us}}^{\mathtt{ODDM}}[\kappa'] &= \!\! \sum_{n=0}^{N\!-\!1} \sum_{l=0}^{M\!-\!1} \sum_{k=0}^{N\!-\!1} D[l,k] p[\kappa'\!-\!lL_{\rm{us}}\!-\!nM'] e^{j \frac{2\pi k(\kappa'-lL_{\rm{us}})}{NM'}}\nonumber \\
&= \! \sum_{n=0}^{N\!-\!1} \sum_{k=0}^{N\!-\!1} \! \left(\sum_{l=0}^{M\!-\!1} \! D[l,k] p_k[(\kappa'\!-\!nM')\!-\!lL_{\rm{us}}] \right) \! e^{j \frac{2 \pi kn}{N}}\nonumber\\
&=\! \sum_{n=0}^{N\!-\!1} \sum_{k=0}^{N\!-\!1} \! \left( D[\frac{\kappa'-nM'}{L_{\rm{us}}},k] * p_k[\kappa'\!-nM'] \right) e^{j \frac{2\pi kn}{N}} \nonumber \\ &=\! \sqrt{N} \sum_{n=0}^{N-1} \mathcal{F}_{N,k}^{-1} \Bigl\{ D_{\rm{e}}[\kappa'-nM',k] * p_k[\kappa'\!-\!nM'] \Bigr\},
\end{align}
where the second line in (\ref{eqn:Disc_DT2}) is obtained by multiplication of the first line by $e^{-j \frac{2 \pi k nM'}{NM'}} e^{j \frac{2 \pi k nM'}{NM'}}=1$ and defining the modulated pulse $p_k[l']\!=\!p[l']e^{j \frac{2 \pi k l'}{M'N}}$. Furthermore, with a simple change of variable $lL_{\rm{us}}$, one may realize that $\sum_{l=0}^{M-1}  D[l,k] p_k[(\kappa'-nM')-lL_{\rm{us}}]$ is linear convolution of the $L_{\rm{us}}$-fold expanded signal $D[l,k]$ along the delay dimension with the modulated pulse.

\begin{figure*}[!t]
  \centering 
  {\includegraphics[scale=0.46]{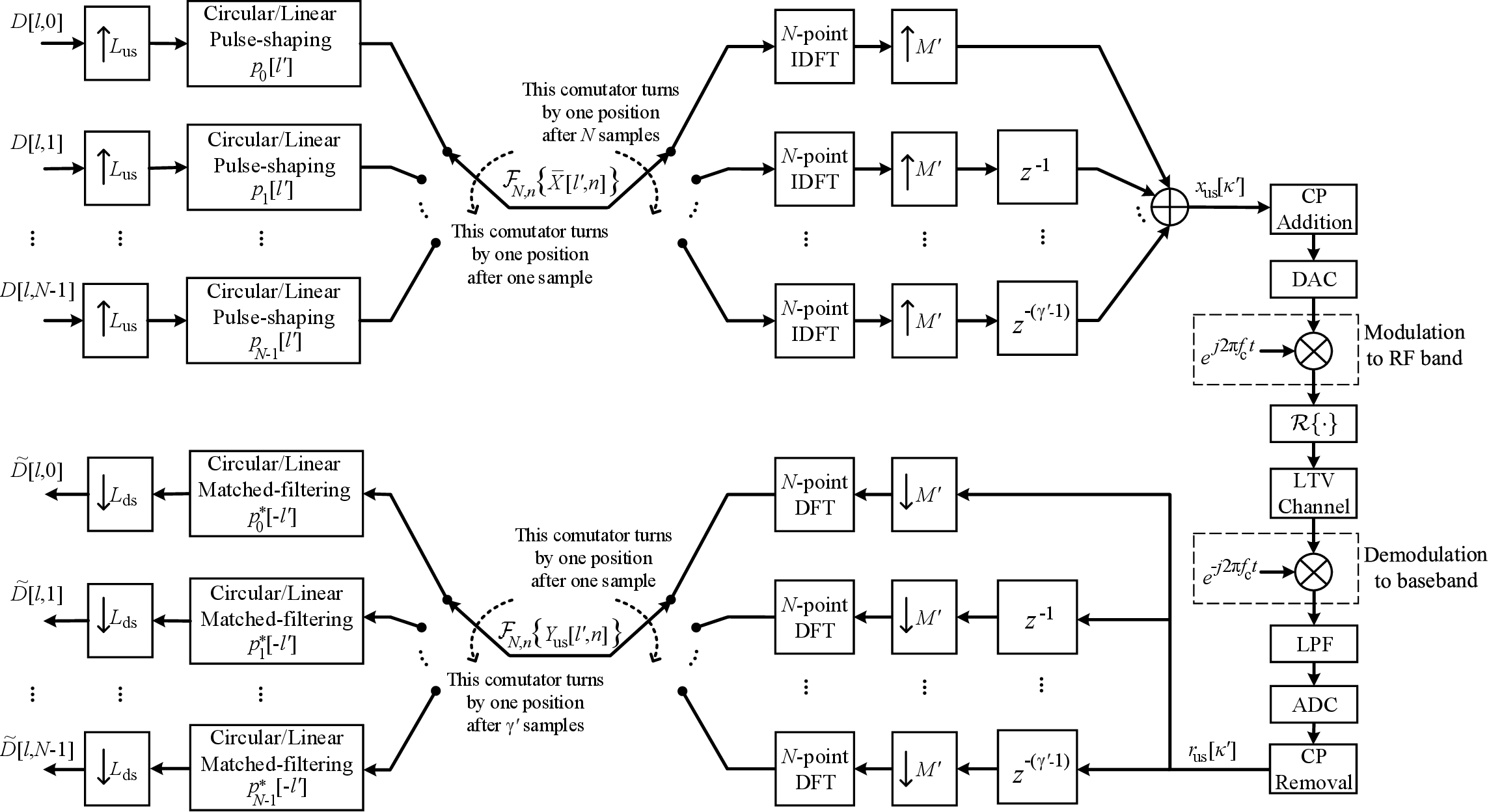}}
  \caption{Generalized modem structure for pulse-shaping on the delay-Doppler plane.}
  \label{fig:imp}
\end{figure*}

In its initial proposal, ODDM was presented as the combination of $M$ shifted and staggered pulse-shaped OFDM signals, $x^{\mathtt{ODDM}}_{l}(t)=\sum_{k=0}^{N-1}D[l,k] e^{j2 \pi \frac{kt}{NT}} u(t)$, with $N$ Doppler subcarriers each, i.e., $x^{\mathtt{ODDM}}(t) = \sum_{l=0}^{M-1} x^{\mathtt{ODDM}}_l(t - l\frac{T}{M})$ \cite{Lin2022_1, Lin2022_2}. In a more recent work, \cite{Tong2023}, ODDM transmit signal was approximated by sample-wise pulse-shaping of the serialized delay-time signal $X[l,n]$ which leads to the same L-PS OTFS signal as in (\ref{eqn:lotfs_p/s3}). However, this approximation is not valid when the condition, $2Q\ll M$, is not satisfied. 

Comparing our derivation in (\ref{eqn:Disc_DT2}) with (\ref{eqn:lotfs_p/s3}), it is clear that ODDM is a linear pulse-shaping method that is totally distinct from L-PS OTFS. The difference between ODDM and L-PS OTFS lies in the pulse-shape choice at different Doppler bins. In ODDM, pulse-shaping along the delay dimension, for a given Doppler bin $k$, is performed by the modulated pulse to the corresponding Doppler frequency, i.e., $p_k[l']$. This important aspect, which is not captured by the formulation in \cite{Lin2022_1}, leads to a staircase behavior in the ODDM spectrum. Our numerical results in Section~\ref{sec:Numerical} also confirm this interesting behavior.
On the contrary, in L-PS OTFS, the same pulse, $p[l']$, is utilized for pulse-shaping along the delay dimension at all Doppler bins, see (\ref{eqn:lotfs_p/s3}).

Similar to L-PS OTFS, the received signal after CP removal is denoted as $r_{\rm{us}}^{\mathtt{ODDM}}[\kappa']$. Then, we rearrange $r_{\rm{us}}^{\mathtt{ODDM}}[\kappa']$ as the 2D delay-time domain signal $Y_{\rm{us}}^{\mathtt{ODDM}}[l',n]=r_{\rm{us}}^{\mathtt{ODDM}}[l'+nM']$ where $l'$ ranges from $-Q'$ to $M'+Q'-1$ and $n=0,\ldots,N-1$. 
Due to the dependence of the pulse-shaping filter on the Doppler index, pulse-shaping at the transmitter and matched filtering at the receiver can only be performed in the delay-Doppler domain. This is another difference between ODDM and L-PS OTFS. Hence, to obtain the received data symbols, we first convert the signal $Y_{\rm{us}}^{\mathtt{ODDM}}[l',n]$ to the delay-Doppler domain. Then, we apply the matched filter, $p^*_k[-l'] \neq p_k[l']$, followed by downsampling along the delay dimension to find
\begin{align} \label{eqn:Do_tilde} \widetilde{D}[l,k] &=  \frac{1}{\sqrt{N}} \Big(\mathcal{F}_{N,n} \Bigl\{  Y_{\rm{us}}^{\mathtt{ODDM}}[l',n] \Bigr\} * p^*_k[-l']\Big)_{l',\downarrow_{L_{\rm ds}}}. \end{align}

\vspace{-0.2cm}
\section{A Unified Framework for Pulse-Shaping on Delay-Doppler Plane}\label{sec:FW}

In this section, we introduce a unified framework and a generalized transceiver structure for pulse-shaping on delay-Doppler plane, based on derivations in Sections~\ref{sec:CPS} and \ref{sec:LPS}. Our generalized framework brings deep insights into the properties, similarities, and differences between the pulse-shaping techniques under study. 

Based on the signal representations in (\ref{eqn:cotfs_p/s4}), (\ref{eqn:lotfs_p/s3}), and (\ref{eqn:Disc_DT2}), all the delay-Doppler plane pulse-shaping techniques under study can be implemented in four stages; (i) $L_{\rm{us}}$-fold expansion of the delay-Doppler domain data symbols along the delay dimension, (ii) pulse-shaping the resulting signal along the delay dimension, (iii) IDFT operation across the Doppler dimension, and (iv) parallel-to-serial conversion by overlap-and-add operation along the delay dimension, i.e., overlapping the delay blocks every $M'$ samples. On this basis, we present all the pulse-shaping techniques in Sections~\ref{sec:CPS} and \ref{sec:LPS} under the following unified framework.
\begin{align} \label{eqn:fw_tx} x_{\rm{us}}[\kappa'] = \sum_{n=0}^{N-1} \! \mathcal{F}_{N,k}^{-1} \Bigl\{ D_{\rm{e}}[\kappa'\!-\!nM',k] \star p_k[\kappa'\!-\!nM'] \Bigr\}, \end{align}
where $\star$ denotes either the circular or linear convolution. In C/L-PS OTFS, the same pulse $p_k[l']=p_0[l']$ is utilized for all the Doppler bins while in ODDM the modulated pulses, $p_k[l']=p[l'] e^{j2 \pi \frac{k l'}{M'N}}$ are utilized for different Doppler bins. After appending a CP to $x_{\rm{us}}[\kappa']$, the resulting signal is transmitted over the wireless channel.

Considering the received delay-Doppler domain symbols in (\ref{eqn:D_tilde}), (\ref{eqn:Dl_tilde}), and (\ref{eqn:Do_tilde}), the receiver for all the pulse-shaping techniques under study can be implemented by (i) converting the received signal after CP removal to the 2D signal $Y_{\rm{us}}[l',n]=r_{\rm{us}}[nM'+l']$, (ii) DFT operation across time dimension, (iii) matched filtering, and (iv) downsampling with the factor $L_{\rm{ds}}$, both along the delay dimension.
Hence, the received symbols in the delay-Doppler domain can be obtained in the following generalized form.
\begin{align} \label{eqn:fw_rx} \widetilde{D}[l,k] = \Big(\mathcal{F}_{N,n} \Bigl\{  Y_{\rm{us}}[l',n] \Bigr\} \star p^*_k[-l']\Big)_{l',\downarrow_{L_{\rm ds}}},
\end{align}
where $p^*_k[-l']=p_0[l'],~\forall k$ for C/L-PS OTFS and $p^*_k[-l']=p[-l'] e^{-j2 \pi \frac{k l'}{M'N}}$ for ODDM. 
Using (\ref{eqn:fw_tx}) and (\ref{eqn:fw_rx}), we present the generalized modem structure for pulse-shaping on delay-Doppler plane in Fig.~\ref{fig:imp},
where $\gamma'$ represents the length of each delay block after pulse-shaping.

\section{Generalized Input-Output Relationships}\label{sec:Input-Output}

In this section, we derive input-output relationships for C/L-PS OTFS and ODDM. To this end, we cast the generalized framework of Section~\ref{sec:FW} into matrix form. Then, considering the underlying operations and matrix structures for each pulse-shaping technique, we derive the effective channel expressions. These expressions reveal how pulse-shaping affects the equivalent delay-Doppler domain channel. Furthermore, our derivations provide insights into the similarities and differences between the pulse-shaping options under investigation.

Considering the aforementioned four pulse-shaping steps in Section~\ref{sec:FW}, and stacking the data symbols $D[l,k]$ into the vector $\bm{d}$, we can represent (\ref{eqn:fw_tx}) in matrix form as
\begin{equation} \label{eqn:x} \bm{x}_{\rm{us}} = \underbrace{\bm{O}}_{\rm{(iv)}} \underbrace{(\bm{F}^{\rm{H}}_{N} \otimes \bm{I}_{\gamma'})}_{\rm{(iii)}}  \underbrace{ \boldsymbol{\Gamma} }_{\rm{(ii)}} \underbrace{(\bm{I}_{N}  \otimes \bm{U})}_{\rm{(i)}} \bm{d}, \end{equation}
where $\bm{d}=[d[0],\ldots,d[{MN-1}]]^{\rm T}$, $d[{l+kM}]=D[l,k]$ for $l=0,\ldots,M-1$, $k=0,\ldots,N-1$. Roman numbers in (\ref{eqn:x}) refer to the pulse-shaping steps.
Multiplication of $\bm{I}_{N}  \otimes \bm{U}$ to $\bm{d}$ realizes the $L_{\rm us}$-fold upsampling operation along the delay dimension. The $M'\times M$ upsampling matrix $\bm{U}$ is formed by inserting $L_{\rm us}-1$ zero rows in between the consecutive rows of $\bm{I}_{M}$, i.e., $\bm{U}=(\bm{I}_M \otimes \bm{z})$, where $\bm{z}=[1,\bm{0}_{1\times (L_{\rm{us}}\!-\!1)}]^{\rm{T}}$.
$\boldsymbol{\Gamma}=(\bm{I}_N \otimes \bm{P}) \odot \bm{E}$ is the pulse-shaping matrix whose size depends on the pulse-shaping technique being deployed, i.e., linear or cyclic.
Hence, for the sake of generality, we consider $\bm{P}$ to be a $\gamma' \times M'$ matrix that is comprised of the pulse-shaping filter coefficients. 
Element-wise multiplication of the matrix $\bm{E}$ into $\bm{I}_N \otimes \bm{P}$ forms the modulated pulses $p_k[l']$ in (\ref{eqn:fw_tx}).
IDFT operation along the Doppler dimension is performed by the matrix $\bm{F}^{\rm{H}}_{N} \otimes \bm{I}_{\gamma'}$.
Lastly, the matrix $\bm{O}=[\bm{O}_0,\ldots,\bm{O}_{N-1}]$ shifts and overlaps the consecutive pulse-shaped delay blocks every $M'$ samples to realize the summation in (\ref{eqn:fw_tx}). Given $\bm{O}_0=[\bm{I}_{\gamma'}^{\rm T},\bm{0}_{(\xi'-\gamma') \times \gamma'}^{\rm T}]^{\rm T}$ where $\xi'$ is length of $\bm{x}_{\rm{us}}$, the sub-matrices $\bm{O}_n$ are formed by circularly shifting the elements on the columns of $\bm{O}_0$ by $nM'$ positions towards the end.

After appending the CP at the beginning of the signal $\bm{x}_{\rm{us}}$, it is transmitted through the channel. Considering perfect synchronization at the receiver side, we first discard the CP from the received signal to obtain
\be \label{eqn:mtx_r} \bm{r}_{\rm{us}} = \bm{R}_{\rm{cp}} \bm{H} \bm{A}_{\rm{cp}} \bm{x}_{\rm{us}} + \boldsymbol{\eta}_{\rm us}, \ee
where $\bm{A}_{\rm{cp}}=[\bm{J}_{\rm{cp}}^{\rm{T}},\bm{I}_{\xi'}^{\rm{T}}]^{\rm{T}}$ and $\bm{R}_{\rm{cp}}=[\bm{0}_{\xi' \times L'_{\rm{cp}}},\bm{I}_{\xi'}]$ are the CP addition and CP removal matrices, respectively. The matrix $\bm{J}_{\rm{cp}}$ is comprised of the last $L'_{\rm{cp}}$ rows of the identity matrix $\bm{I}_{\xi'}$ and $\bm{r}_{\rm{us}} =[r_{\rm{us}}[0],\ldots,r_{\rm{us}}[\xi'-1]]^{\rm T}$. The time-varying channel, that is oversampled by the factor of $L_{\rm us}$, is represented by the matrix $\bm{H}$, with the elements $[\bm{H}]_{\kappa',i}=h_{\rm us}[\kappa',\kappa'-i]$ for $\kappa'=0,\ldots,\xi'+L'_{\rm{cp}}-1$ and $\boldsymbol{\eta}_{\rm us}=[\eta_{\rm us}[0],\ldots,\eta_{\rm us}[\xi'-1]]^{\rm T}$ is the noise vector in delay-time domain. 

By applying the reverse procedure to the one at the transmitter in (\ref{eqn:x}), we can rearrange (\ref{eqn:fw_rx}) in matrix form as
\begin{equation} \label{eqn:mtx_d_hat} \widetilde{\bm{d}} = (\bm{I}_N \otimes \bm{U}^{\rm{H}})  {\boldsymbol{\Gamma}}^{\rm{H}} (\bm{F}_N \otimes \bm{I}_{\gamma'})  \bm{O}^{\rm{H}}  \bm{r}_{\rm{us}}, \end{equation}
where $\bm{O}^{\rm{H}}$ collects and concatenates the delay blocks of the length $\gamma'$ from the received signal and $\bm{F}_N \otimes \bm{I}_{\gamma'}$ takes them to the delay-Doppler domain. Then, ${\boldsymbol{\Gamma}}^{\rm{H}}$ performs matched filtering, and $\bm{I}_N \otimes \bm{U}^{\rm{H}}$ downsamples the resulting signal to obtain the received delay-Doppler domain symbols in the vector $\widetilde{\bm{d}}=[\widetilde{d}[0],\ldots,\widetilde{d}[MN-1]]^{\rm T}$ where $\widetilde{d}[l+kM]=\widetilde{D}[l,k]$ for $l=0,\ldots,M-1$, $k=0,\ldots,N-1$.

By substituting (\ref{eqn:x}) and (\ref{eqn:mtx_r}) into (\ref{eqn:mtx_d_hat}), we arrive at the following generalized input-output relationship for all the pulse-shaping techniques that were discussed earlier.
\begin{align} \label{eqn:mtx_d_hat1} \widetilde{\bm{d}}\!&=\!\!\Big[ \boldsymbol{\Gamma}_{\rm{ds}} (\bm{F}_N \!\otimes\! \bm{I}_{\gamma'}\!) \bm{O}^{\rm{H}} \boldsymbol{\mathcal{H}} \bm{O} (\bm{F}^{\rm{H}}_{N} \!\otimes\! \bm{I}_{\gamma'}\!) \boldsymbol{\Gamma}_{\rm{us}} \Big] \bm{d} + \widetilde{\boldsymbol{\eta}}_{\rm{us}} \nonumber \\ &= \bm{H}_{\rm{eff}} \bm{d} + \widetilde{\boldsymbol{\eta}}_{\rm{us}}, \end{align}
where $\boldsymbol{\Gamma}_{\rm{us}} = \boldsymbol{\Gamma} (\bm{I}_{N} \otimes \bm{U}) = \big((\bm{I}_N \otimes \bm{P}) \odot \bm{E} \big)(\bm{I}_{N} \otimes \bm{U})$, $\boldsymbol{\Gamma}_{\rm{ds}} = \boldsymbol{\Gamma}_{\rm{us}}^{\rm{H}}$, $\boldsymbol{\mathcal{H}}=\bm{R}_{\rm{cp}} \bm{H} \bm{A}_{\rm{cp}}$ and $\widetilde{\boldsymbol{\eta}}_{\rm{us}}=\boldsymbol{\Gamma}_{\rm{ds}} (\bm{F}_N \otimes \bm{I}_{\gamma'}) \bm{O}^{\rm{H}} \bm{R}_{\rm{cp}} \boldsymbol{\eta}_{\rm{us}}$ is the delay-Doppler domain noise vector. Based on (\ref{eqn:mtx_d_hat1}), $\bm{H}_{\rm{eff}}= \boldsymbol{\Gamma}_{\rm{ds}} (\bm{F}_N \otimes \bm{I}_{\gamma'}) \bm{O}^{\rm{H}} \boldsymbol{\mathcal{H}} \bm{O} (\bm{F}^{\rm{H}}_{N} \otimes \bm{I}_{\gamma'}\!) \boldsymbol{\Gamma}_{\rm{us}}$ is the end-to-end channel matrix that captures the effect of pulse-shaping in delay-Doppler domain. Hence, in the following subsections, we take a deep dive into the pulse-shaping effect on the equivalent channel for different techniques.

\vspace{-0.3cm}
\subsection{Cyclic/Circular Pulse-Shaping} \label{subsec:ch_cotfs}
Let us recall that in C-PS OTFS, the same pulse-shape is used for all the Doppler bins, and pulse-shaping is performed by circular convolution. Therefore, the matrix $\bm{P}$ in $\boldsymbol{\Gamma}$ is circulant, i.e., $\bm{P}\!\!=\!\!\bm{P}_\mathtt{C}\!\!=\!\!{\rm{circ}}\{ \bm{p} \}$, where the subscript $\mathtt{C}$ is used to emphasize cyclic pulse-shaping and the vector $\bm{p}=[p[0],p[1],\ldots,p[M'-1]]^{\rm{T}}$ includes the pulse-shaping filter coefficients. Consequently, the pulse-shaping matrix is represented as $\boldsymbol{\Gamma}_{\rm{us}}\!=\!\boldsymbol{\Gamma}_{\mathtt{C}}\!=\!(\bm{I}_N \!\otimes\! \bm{P}_{\mathtt{C}})(\bm{I}_{N} \!\otimes\! \bm{U})=(\bm{I}_N \!\otimes\! (\bm{P}_{\mathtt{C}} \bm{U}))$. Since, after cyclic pulse-shaping, the length of each time slot remains the same, i.e., $\gamma'=M'$, there is no overlap between the adjacent delay blocks. Thus, $\bm{O}=\bm{I}_{M'N}$, and
the effective channel boils down to
\begin{align} \label{eqn:ch_c1} \bm{H}_{\rm{eff}}^\mathtt{C} &= \boldsymbol{\Gamma}_{\mathtt{C}}^{\rm{H}} (\bm{F}_{N} \otimes \bm{I}_{M'}) \boldsymbol{\mathcal{H}} (\bm{F}^{\rm{H}}_{N} \otimes \bm{I}_{M'}) \boldsymbol{\Gamma}_{\mathtt{C}} \nonumber \\ &= (\bm{F}_N \otimes \bm{I}_{M}) (\boldsymbol{\Gamma}_{\mathtt{C}}^{\rm{H}} {\boldsymbol{\mathcal{H}}} \boldsymbol{\Gamma}_{\mathtt{C}}) (\bm{F}^{\rm{H}}_N \otimes \bm{I}_{M}). \end{align}
The first line in (\ref{eqn:ch_c1}) represents the equivalent channel in the delay-Doppler domain. By substituting $\boldsymbol{\Gamma}_{\mathtt{C}}=(\bm{I}_N \!\otimes\! (\bm{P}_{\mathtt{C}} \bm{U}))$ in the first line of (\ref{eqn:ch_c1}) and using Kronecker product properties, we obtain the equivalent channel in the delay-time domain in the second line, i.e., $\boldsymbol{\Gamma}_{\mathtt{C}}^{\rm{H}} {\boldsymbol{\mathcal{H}}} \boldsymbol{\Gamma}_{\mathtt{C}}$.

It is also worth investigating the effective channel when our proposed OOB reduction techniques in Sections~\ref{subsec:OOB reduction} and \ref{subsec:OOB reduction ZP} are deployed. In both cases, the number of delay bins is increased from $M$ to $M_{\rm{CE}}= M+L_{\rm{CE}}$, assuming $L_{\rm{ZG}}=L_{\rm{CE}}$. This increases the length of $\bm{d}$ in (\ref{eqn:mtx_d_hat1}) to $M_{\rm{CE}}N$. In the windowing-based technique of Section~\ref{subsec:OOB reduction}, the transmit data symbols are cyclically extended at both sides along the delay dimension to facilitate the windowing operation. Consequently, the effective channel for C-PS OTFS with windowing can be obtained as 
\begin{align} \label{eqn:CW} \bm{H}^{\mathtt{CW}}_{\rm{eff}} &= \boldsymbol{\Gamma}_{\mathtt{CW}}^{\rm{H}} (\bm{F}_{N} \! \otimes \! \bm{I}_{M'_{\rm{CE}}}) \boldsymbol{\mathcal{H}} (\bm{F}^{\rm{H}}_{N} \! \otimes \! \bm{I}_{M'_{\rm{CE}}}) (\bm{I}_N \! \otimes \! \bm{W}) \boldsymbol{\Gamma}_{\mathtt{CW}} \nonumber \\ &= (\bm{F}_{N} \! \otimes \! \bm{I}_{M_{\rm{CE}}}) \boldsymbol{\Gamma}_{\mathtt{CW}}^{\rm{H}} \boldsymbol{\mathcal{H}} (\bm{I}_N \! \otimes \! \bm{W}) \boldsymbol{\Gamma}_{\mathtt{CW}} (\bm{F}^{\rm{H}}_{N} \! \otimes \! \bm{I}_{M_{\rm{CE}}}), \end{align}
where $\bm{I}_N \otimes \bm{W}$ realizes the windowing operation along the delay dimension and $M'_{\rm{CE}}=M_{\rm{CE}}L_{\rm{us}}$. $\bm{W}=\rm{diag}\{\boldsymbol{\psi}_{\beta, \rm{w}}\}$ is the windowing matrix that includes the RRC window coefficients on its diagonal elements, i.e., $\boldsymbol{\psi}_{\beta,\rm{w}}\!=\![\psi_{\beta,\rm{w}}[0],\!\ldots\!,\psi_{\beta,\rm{w}}[M'_{\rm{CE}}\!-\!1]]^{\rm T}$. $\boldsymbol{\Gamma}_{\rm{us}}\!=\!\boldsymbol{\Gamma}_{\mathtt{CW}}\!=\!(\bm{I}_N \! \otimes \! (\bm{P}_{\mathtt{CW}} \bm{U}_{\mathtt{CW}}))$ is the pulse-shaping matrix where $\bm{U}_{\mathtt{CW}}=(\bm{I}_{M_{\rm{CE}}} \otimes \bm{z})$, $\bm{P}_{\mathtt{CW}}={\rm{circ}}\{\bm{p_{\mathtt{CW}}}\}$, and $\bm{p}_{\mathtt{CW}}=[p[0],\ldots,p[M'_{\rm{CE}}-1]]^{\rm{T}}$. Equation (\ref{eqn:CW}) is very similar to (\ref{eqn:ch_c1}) with the only difference being the length of the delay dimension and presence of the windowing matrix.

For our proposed OOB reduction technique using ZGs in Section~\ref{subsec:OOB reduction ZP}, given $L_{\rm{ZG}}=L_{\rm{CE}}$, the effective channel is essentially the same as (\ref{eqn:ch_c1}) where $M$ and $M'$ are replaced by $M_{\rm{CE}}$ and $M'_{\rm{CE}}$, respectively.


\vspace{-0.3cm}
\subsection{Linear Pulse-Shaping} \label{sec:ch_lotfs}
As discussed earlier, in L-PS OTFS, the same pulse-shape is deployed for all the Doppler bins and pulse-shaping is performed by linear convolution. 
Considering the truncated pulse after $Q$ zero-crossings at each side of its main lobe, $\bm{P}=\bm{P}_\mathtt{L}=\mathcal{T}_{\gamma' \times M'} \{ \bm{p}^{\mathtt{Trc.}} \}$, where $\gamma'=M'+2Q'$ and $\bm{p}^{\mathtt{Trc.}}=[p[0],p[1],\ldots,p[2Q'-1],\bm{0}_{1 \times M'}]^{\rm{T}}$. Thus, the pulse-shaping matrix is represented as $\boldsymbol{\Gamma}_{\mathtt{L}} = \big(\bm{I}_N \otimes (\bm{P}_{\mathtt{L}} \bm{U}) \big)$ where the subscript $\mathtt{L}$ refers to the linear pulse-shaping. From (\ref{eqn:mtx_d_hat1}), 
the effective channel is obtained as
\begin{align} \label{eqn:ch_l} \bm{H}_{\rm{eff}}^\mathtt{L} &= \boldsymbol{\Gamma}_{\mathtt{L}}^{\rm{H}} (\bm{F}_{N} \otimes \bm{I}_{\gamma'})  \bm{O}^{\rm{H}} \boldsymbol{\mathcal{H}} \bm{O} (\bm{F}^{\rm{H}}_{N} \otimes \bm{I}_{\gamma'}) \boldsymbol{\Gamma}_{\mathtt{L}} \nonumber \\ &= (\bm{F}_N \!\otimes\! \bm{I}_{M}) (\boldsymbol{\Gamma}_{\mathtt{L}}^{\rm{H}} \bm{O}^{\rm{H}} \boldsymbol{\mathcal{H}} \bm{O} \boldsymbol{\Gamma}_{\mathtt{L}}) (\bm{F}^{\rm{H}}_N \otimes \bm{I}_{M}). \end{align}
Following the same line of derivation as in (\ref{eqn:ch_c1}), the equivalent delay-time domain channel is obtained as $\boldsymbol{\Gamma}_{\mathtt{L}}^{\rm{H}} \bm{O}^{\rm{H}} \boldsymbol{\mathcal{H}} \bm{O} \boldsymbol{\Gamma}_{\mathtt{L}}$. 

\begin{figure*}[ht]
  \centering
  {\includegraphics[scale=0.125]{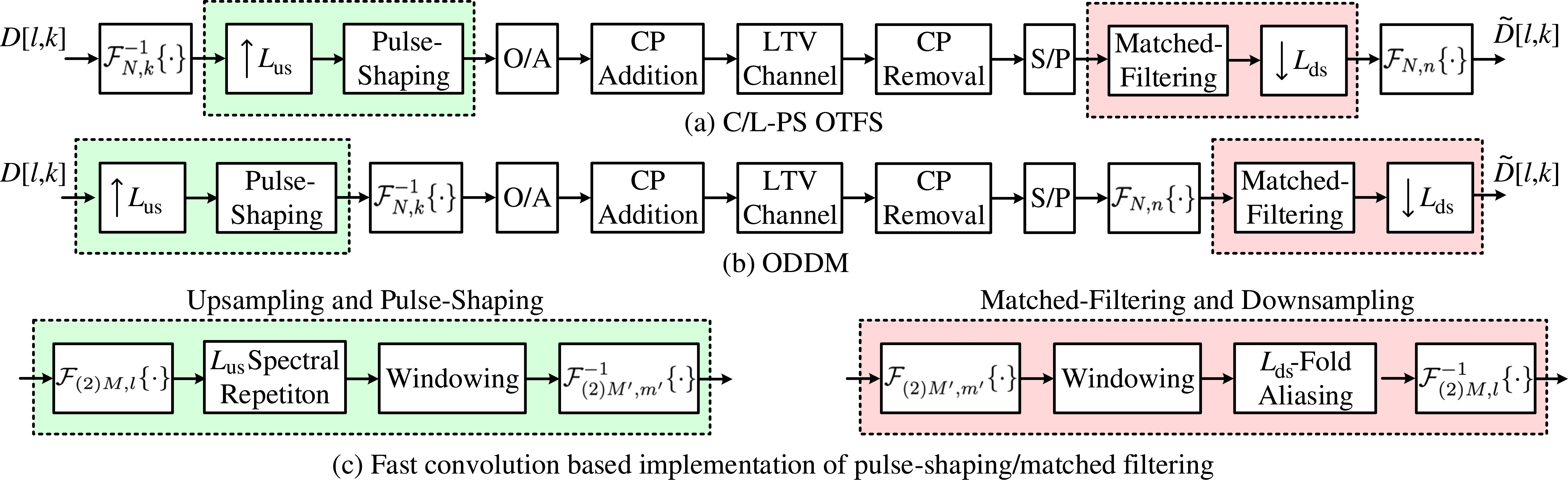}}
  \vspace{-0.4cm}
  \caption{The simplified modem structures.}
  \vspace{-0.5cm}
  \label{fig:lotfs}
\end{figure*}

\vspace{-0.3cm}
\subsection{ODDM} \label{sec:ch_oddm}
As mentioned in Section~\ref{subsection:oddm}, ODDM is a type of linear pulse-shaping where modulated pulses are used at different Doppler bins. Therefore, the equivalent delay-Doppler domain channel for ODDM is similar to that of L-PS OTFS, with the only difference in the pulse-shaping filter. Specifically, the additional exponential term in ODDM pulse-shape leads to $\boldsymbol{\Gamma}_{\rm{\mathtt{ODDM}}} \!\!=\!\! \big( (\bm{I}_N \otimes \bm{P}_\mathtt{L}) \odot \bm{E} \big) (\bm{I}_N \otimes \bm{U}) \!=\! \big( \bm{I}_N \otimes (\bm{P}_\mathtt{L} \bm{U}) \big) \odot \overline{\bm{E}} \!=\! \boldsymbol{\Gamma}_{\mathtt{L}} \odot \overline{\bm{E}}$, where $\overline{\bm{E}}=\bm{E} (\bm{I}_N \otimes \bm{U})$. The matrix $\bm{E}={\rm{diag}}\{ \bm{E}_0, \ldots, \bm{E}_{N-1} \}$, where $\bm{E}_k = \mathcal{T}_{\gamma' \times M'} [\bm{e}_k]$, $\gamma'=M'+2Q'$ and $\bm{e}_k=[1,e^{j \frac{2 \pi k}{M'N}},\ldots,e^{j \frac{2 \pi k(\gamma'-1)}{M'N}}]^{\rm{T}}$ for $k=0,\ldots,N-1$.
Thus, the effective channel in the delay-Doppler domain can be represented as
\begin{align} \label{eqn:ch_ODDM} \!\!\bm{H}_{\rm{eff}}^{\mathtt{ODDM}} \!&=\! (\boldsymbol{\Gamma}_{\mathtt{ODDM}}^{\rm{DD}})^{\rm{H}} (\bm{F}_N \! \otimes \! \bm{I}_{\gamma'}) \bm{O}^{\rm{H}} \boldsymbol{\mathcal{H}} \bm{O} (\bm{F}^{\rm{H}}_{N} \! \otimes \! \bm{I}_{\gamma'}) \boldsymbol{\Gamma}_{\mathtt{ODDM}}^{\rm{DD}} \nonumber \\
\!&=\! (\bm{F}_N \otimes \bm{I}_M) (\bm{\Gamma}^{\rm{DT}}_{\mathtt{ODDM}})^{\rm{H}} \bm{O}^{\rm{H}} \boldsymbol{\mathcal{H}} \bm{O} \bm{\Gamma}^{\rm{DT}}_{\mathtt{ODDM}} (\bm{F}_{N}^{\rm{H}} \otimes \bm{I}_M), \end{align}
where 
$\bm{\Gamma}^{\rm{DT}}_{\mathtt{ODDM}}=(\bm{F}_N^{\rm{H}} \otimes \bm{I}_{\gamma'})(\boldsymbol{\Gamma}_{\mathtt{L}} \odot \overline{\bm{E}})(\bm{F}_N \otimes \bm{I}_M)$ and $\bm{\Gamma}^{\rm{DD}}_{\mathtt{ODDM}}\!=\!\boldsymbol{\Gamma}_{\mathtt{L}} \odot \overline{\bm{E}}$ are the pulse-shaping matrices in the delay-Doppler and delay-time domains, respectively. Hence, the equivalent delay-time domain channel is obtained as $(\bm{\Gamma}^{\rm{DT}}_{\mathtt{ODDM}})^{\rm{H}} \bm{O}^{\rm{H}} \boldsymbol{\mathcal{H}} \bm{O} \boldsymbol{\Gamma}^{\rm{DT}}_{\mathtt{ODDM}}$.

From equations (\ref{eqn:ch_c1}) to (\ref{eqn:ch_ODDM}), one may realize that the transmit and receive pulse-shapes are absorbed into the effective channel. Hence, the choice of pulse-shape contributes to the effective channel length. This highlights the importance of utilizing short transmit and receive pulse-shaping filters.

\vspace{-0.2cm}
\section{Low Complexity Implementation}\label{sec:Comp}

Using the unified framework in Section~\ref{sec:FW}, in this section, we propose low-complexity implementation of the pulse-shaping techniques under study. We also analyze and compare the computational load of our proposed techniques in terms of the number of complex multiplications (CMs) with their direct implementation and the one in \cite{Lin2022_2} which is summarized in Table~\ref{Table:Complexity}. 
Since the receiver operations are dual of those at the transmitter, our focus in the following is on the modulator. 

The IDFT operation in the generalized modulator expression, (\ref{eqn:fw_tx}), can be efficiently implemented using the inverse fast Fourier transform (IFFT) algorithm. However, direct implementation of the convolution operation in (\ref{eqn:fw_tx}) is computationally complex.
For the reasons that were discussed in Section~\ref{sec:LPS}, the pulse $p[l']$ may be truncated. This reduces the convolution complexity. Nevertheless, perfect reconstruction of the transmitted data symbols at the receiver cannot be achieved when the pulse is truncated. Hence, to reduce the computational load without relying on pulse truncation, in the following, we propose to implement pulse-shaping in the frequency domain using the modem structures in Fig.~\ref{fig:lotfs}. 
 
Due to the use of a Doppler independent pulse in C/L-PS OTFS, it is more computationally efficient to perform pulse-shaping in the delay-time domain than in the delay-Doppler domain. This involves $M$ number of $N$-point IFFT operations along the Doppler dimension which requires $\frac{MN}{2}\log_2N$ CMs. 
Direct implementation of convolution operation that is required for pulse-shaping demands $\frac{NM}{4}(2Q'-2Q)$ and $\frac{NM}{4}(M'-M)$ CMs with and without pulse truncation, respectively. This involves scaling and adding the delayed pulse by each data symbol, \cite{Lin2022_2}. The factor of $\frac{1}{4}$ is due to the real and symmetric properties of the pulse and subtraction of $2Q$ and $M$ are to avoid multiplications at the zero-crossing instants.

Based on our derivations in Section~\ref{subsec:CPS-OTFS}, cyclic pulse-shaping can be efficiently implemented in the frequency domain by a simple multiplication operation, see equation (\ref{eqn:X_bar}) and Figs.~\ref{fig:lotfs}~(a) and (c). To move from the delay-time domain to the time-frequency domain and vice versa, $N$ number of $M$-point fast Fourier transform (FFT) and IFFT operations, respectively, with $2\times\frac{MN}{2}\log_2M$ CMs in total are required. Considering the pulse-shaping filter with real-valued coefficients in (\ref{eqn:rrc_freq}), multiplication is only required during the roll-off period that includes $\lceil \alpha M\rceil$ samples. Thus, frequency domain pulse-shaping requires $\frac{N}{2}\lceil \alpha M\rceil$ CMs.

\renewcommand{\arraystretch}{1.4}
\begin{table}
\vspace{-0.2cm}
\centering
\caption{Implementation Complexity}\vspace{-0.2cm}
\resizebox{0.489\textwidth}{!}
{\begin{tabular}{| c || c |}
\hline\hline
\textbf{Pulse-shaping Technique}
& \textbf{Number of CMs at the Modulator/Demodulator}
\\ \hline \hline
Direct C/L-PS OTFS Implementation
& $NM (\frac{1}{2} \log_2N + \frac{1}{4}(2Q' - 2Q))$~ for ~$Q\leq M/2$
\\ \hline
ODDM Implementation in \cite{Lin2022_2}
& $N^2M (2Q' - 2Q)$
\\ \hline
Proposed C-PS OTFS Implementation
& $NM (\frac{1}{2} \log_2{N} + \frac{1}{2} \log_2{M} + \frac{L_{\rm{us}}}{2} \log_2{M'}) + \frac{N}{2} \lceil \alpha M \rceil$
\\ \hline
Proposed L-PS OTFS Implementation
& $NM(\frac{1}{2} \log_2{N} + \log_2{2M}+ L_{\rm{us}}\log_2{2M'}) + \frac{N}{2} \lceil 2 \alpha M \rceil$
\\ \hline
Proposed ODDM Implementation
& $NM(\log_2{2M} + L_{\rm{us}} \log_2{2NM'}) + N \lceil 2 \alpha M \rceil$
\\ \hline\hline
\end{tabular}
\label{Table:Complexity}
}
\vspace{-0.6cm}
\end{table}

To efficiently implement the linear convolution operation of L-PS OTFS in the frequency domain, we take the fast convolution approach \cite{proakis2007digital}. Thus, we zero-pad each delay block so that its length becomes at least equal to $\gamma'$. Since, $M$ is usually chosen as powers of two, we pad the delay blocks with $M$ zeros for an efficient implementation of FFT/IFFT. Consequently, as shown in Fig.~\ref{fig:lotfs}, linear convolution of L-PS OTFS can be performed similar to C-PS OTFS. Compared to C-PS OTFS, in L-PS OTFS, $2M$ instead of $M$ samples at each time-slot in delay/frequency domain need to be processed. 

Due to its Doppler dependent pulse, the delay-time domain pulse-shaping does not suit ODDM. To simplify ODDM implementation, the authors in \cite{Lin2023_2} have approximated ODDM with L-PS OTFS,  under the condition $2Q\ll M$. However, such approximation may not be very accurate when this condition is not satisfied. To the best of our knowledge, the pulse-shaped OFDM based structure of ODDM in \cite{Lin2022_2} is the simplest implementation of this waveform without approximation in the existing literature. Using (\ref{eqn:Disc_DT1}), one may realize that ODDM implementation in \cite{Lin2022_2} involves scaling and addition of the modulated and delayed pulse, $u[\kappa']$,  by data symbols. Since the pulse has $N(2Q' - 2Q)$ non-zero samples, ODDM is implemented by $N^2M(2Q' - 2Q)$ CMs. 

To simplify ODDM implementation, recalling (\ref{eqn:fw_tx}), we propose to perform delay-Doppler domain pulse-shaping first with the same fast convolution approach to L-PS OTFS, see Figs.~\ref{fig:lotfs}~(b) and (c). This process requires $N$ number of $2M$-point FFT operations, $N\lceil 2\alpha M\rceil$ CMs in the frequency domain and  $2M'$-point IFFT operations on $N$ time slots. It is worth noting that the frequency responses of the modulated pulses in (\ref{eqn:fw_tx}) are assumed to be stored in the memory. Finally, we translate the pulse-shaped signal to the delay-time domain by $2M'$ number of $N$-point IFFT operations.

\begin{figure}[!t]
  \centering 
  {\includegraphics[scale=0.3]{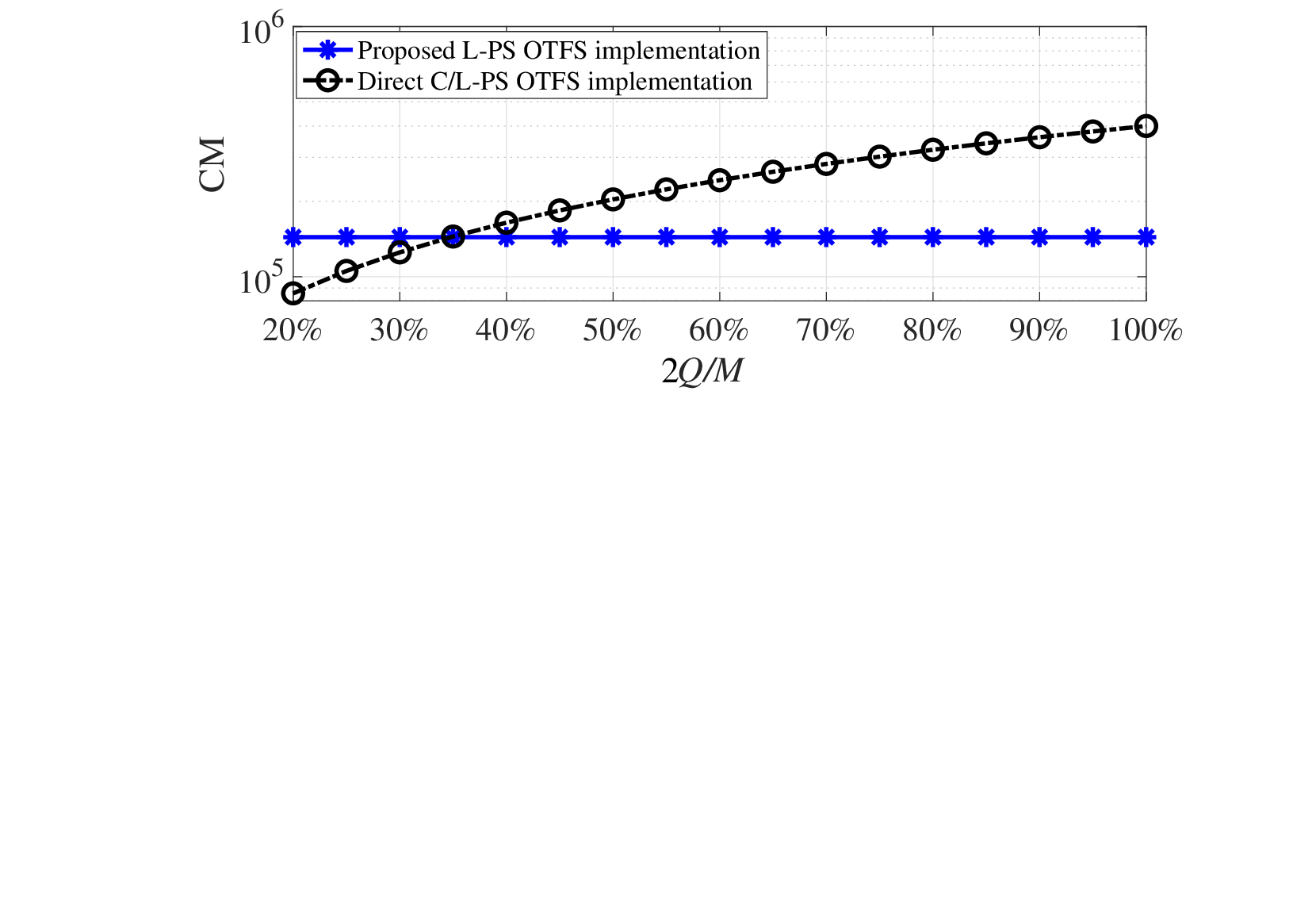}}
  \vspace{-3.6cm}
  \caption{The effect of pulse truncation on the computational complexity of L-PS OTFS implementation for $M=128$, $N=32$, and $L_{\rm{us}}=4$.}
  \label{fig:Comp_Comp}
\end{figure}

\begin{figure}[!t]
  \centering 
  {\includegraphics[scale=0.3]{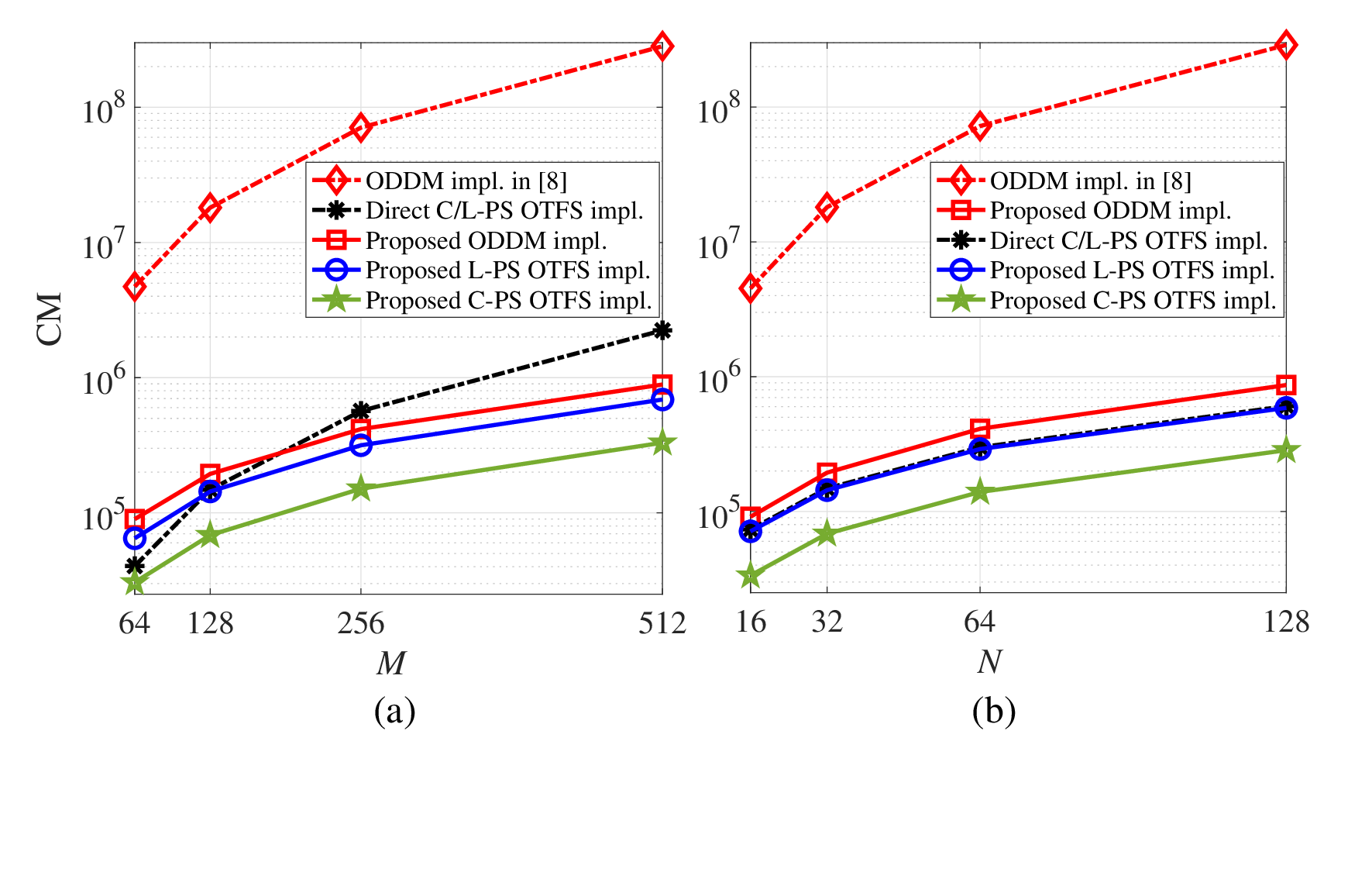}}
  \vspace{-1.5cm}
  \caption{Complexity of different pulse-shaping techniques for $L_{\rm{us}}=4$, $2Q/M=35\%$, and (a) $N=32$, and (b) $M=128$.}
  \vspace{-0.4cm}
  \label{fig:Comp_MN}
\end{figure}

In Figs.~\ref{fig:Comp_Comp} and \ref{fig:Comp_MN}, we quantify and compare the complexity expressions presented in Table~\ref{Table:Complexity}. 
In Fig.~\ref{fig:Comp_Comp}, we investigate the pulse truncation effect on the computational complexity versus the relative truncation level to the delay block length, i.e., $2Q/M$. From Fig.~\ref{fig:Comp_Comp}, when $2Q/M\leq 35\%$, direct implementation of pulse-shaping operation is more computationally efficient than our proposed implementation. However, as $2Q/M$ increases, the advantages of our proposed technique become more evident.
In Fig.~\ref{fig:Comp_Comp}, we only consider L-PS OTFS as our proposed C-PS OTFS implementation has a lower complexity than the direct implementation for all truncation levels. 
Setting $2Q/M = 35\%$, in Fig.~\ref{fig:Comp_MN}, we compare the computational load of our proposed implementation techniques with the one in \cite{Lin2022_2} and their direct implementation.
As it is evident in Fig.~\ref{fig:Comp_MN}, utilizing our proposed techniques, substantial amount of complexity reduction can be achieved for different pulse-shaping techniques.
Particularly, our proposed implementation of the ODDM modulator/demodulator leads to over two orders of magnitude complexity reduction compared to the one in \cite{Lin2022_2}. Furthermore, Fig.~\ref{fig:Comp_MN}~(a) shows that as $M$ increases when $2Q/M$ and $N$ are fixed, direct implementation of pulse-shaping for L-PS OTFS rapidly becomes inefficient. However, increasing $N$ when $2Q/M$ and $M$ are fixed, does not affect efficiency of direct pulse-shaping implementation, see Fig.~\ref{fig:Comp_MN}~(b).

\vspace{-0.2cm}
\section{numerical analysis}\label{sec:Numerical}

In this section, we numerically analyze and compare the performance of different pulse-shaping techniques in terms of OOB emissions, BER, and PAPR. Additionally, we evaluate the efficacy of our proposed OOB reduction techniques that simultaneously improve the BER performance. In our simulations, we consider a delay-Doppler grid with $M=128$ delay bins and $N=32$ Doppler bins. As explained in Section~\ref{sec:Principle}, we append a CP at the beginning of each block that is longer than the channel delay spread. We deploy the SRRC pulse-shaping filter with the roll-off factor $\alpha=0.1$, that is truncated after $Q=12$ zero-crossings, unless otherwise stated. We set the upsampling factor to $L_{\rm{us}}=4$, the carrier frequency to $f_{\rm{c}}\!=\!5.9$~GHz, and the transmission bandwidth as ${\rm BW}=1.92$~MHz.
We use the extended vehicular~A (EVA) channel model \cite{3gpp}, with the maximum Doppler spread $\nu_{\rm{max}}=\frac{v}{c}f_{\rm{c}}$ where the relative velocity of $v=500$~km/h is considered between the transmitter and receiver, and $c$ is the speed of light. Finally, we assume perfect synchronization and perfectly known channel at the receiver where the minimum mean square error (MMSE) equalizer is deployed for detection. 

\begin{figure}[!t]
  \centering 
  {\includegraphics[scale=0.33]{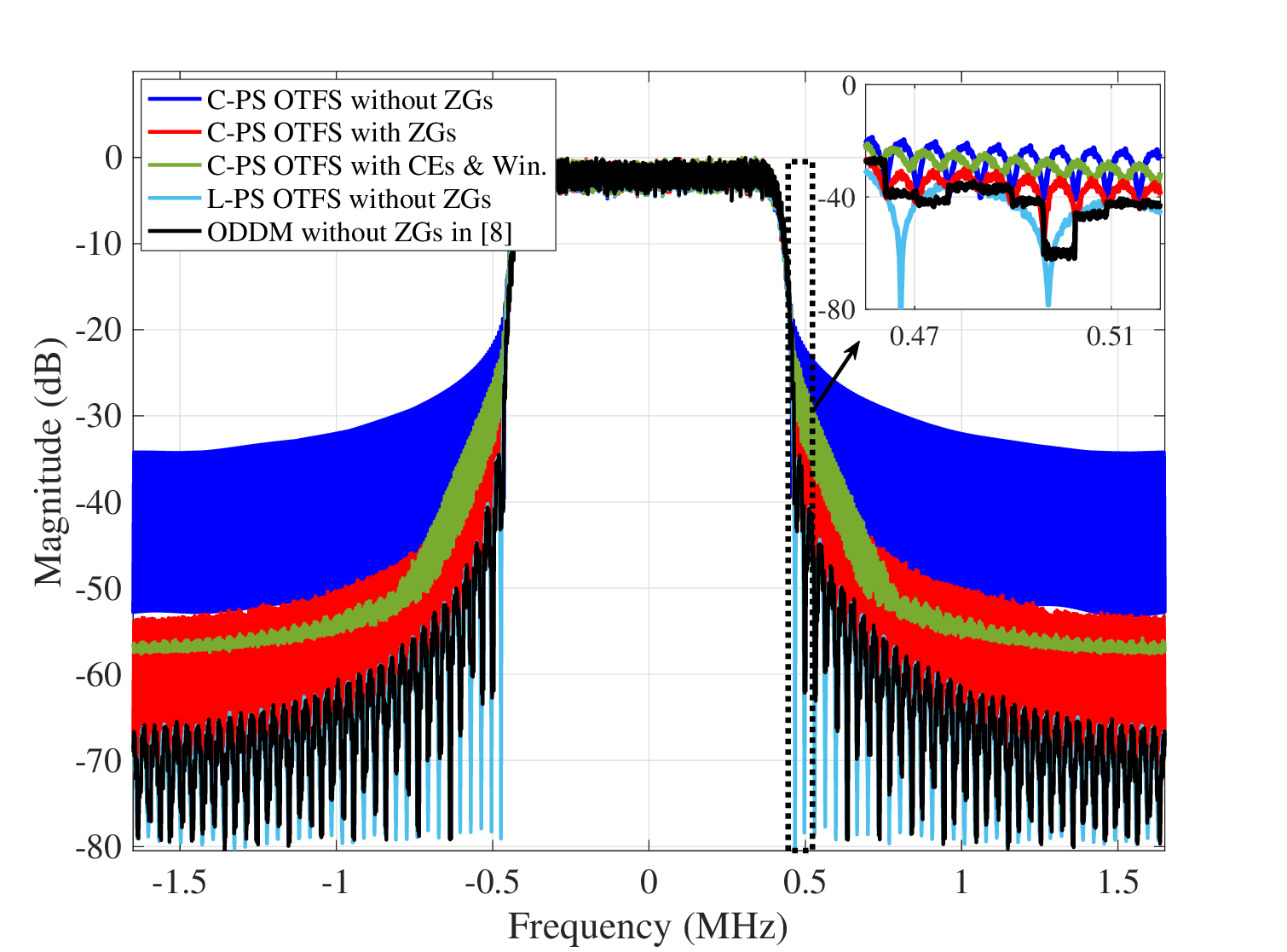}}
  \caption{PSD comparison of different pulse-shaping techniques for $M=128$, $N=32$, $L_{\rm{us}}=4$, and $Q=12$.}
  \label{fig:PSD}
    \vspace{-0.3cm}
\end{figure}

Fig.~\ref{fig:PSD}, compares the OOB emissions of different delay-Doppler multiplexing techniques. According to the discussions in Section~\ref{sec:CPS}, and our results here, C-PS OTFS has large OOB emissions which are effectively suppressed by our proposed solutions using windowing and zero-guards. As shown in Fig.~\ref{fig:PSD}, by investing only $L_{\rm CE}=L_{\rm ZG}=6$ additional samples along the delay dimension to accommodate windowing or ZGs, i.e., $4.47\%$ of the delay block length, a substantial amount of OOB reduction is achieved.  
From the results in Fig.~\ref{fig:PSD}, linear pulse-shaping techniques lead to the lowest OOB emissions among different techniques. An interesting observation here is the staircase behaviour of the ODDM spectrum, which was not reported in the literature previously. This behavior is mathematically justified by the modulated pulse to different Doppler frequencies in (\ref{eqn:Disc_DT2}).

\begin{figure}[!t]
  \centering 
  {\includegraphics[scale=0.33]{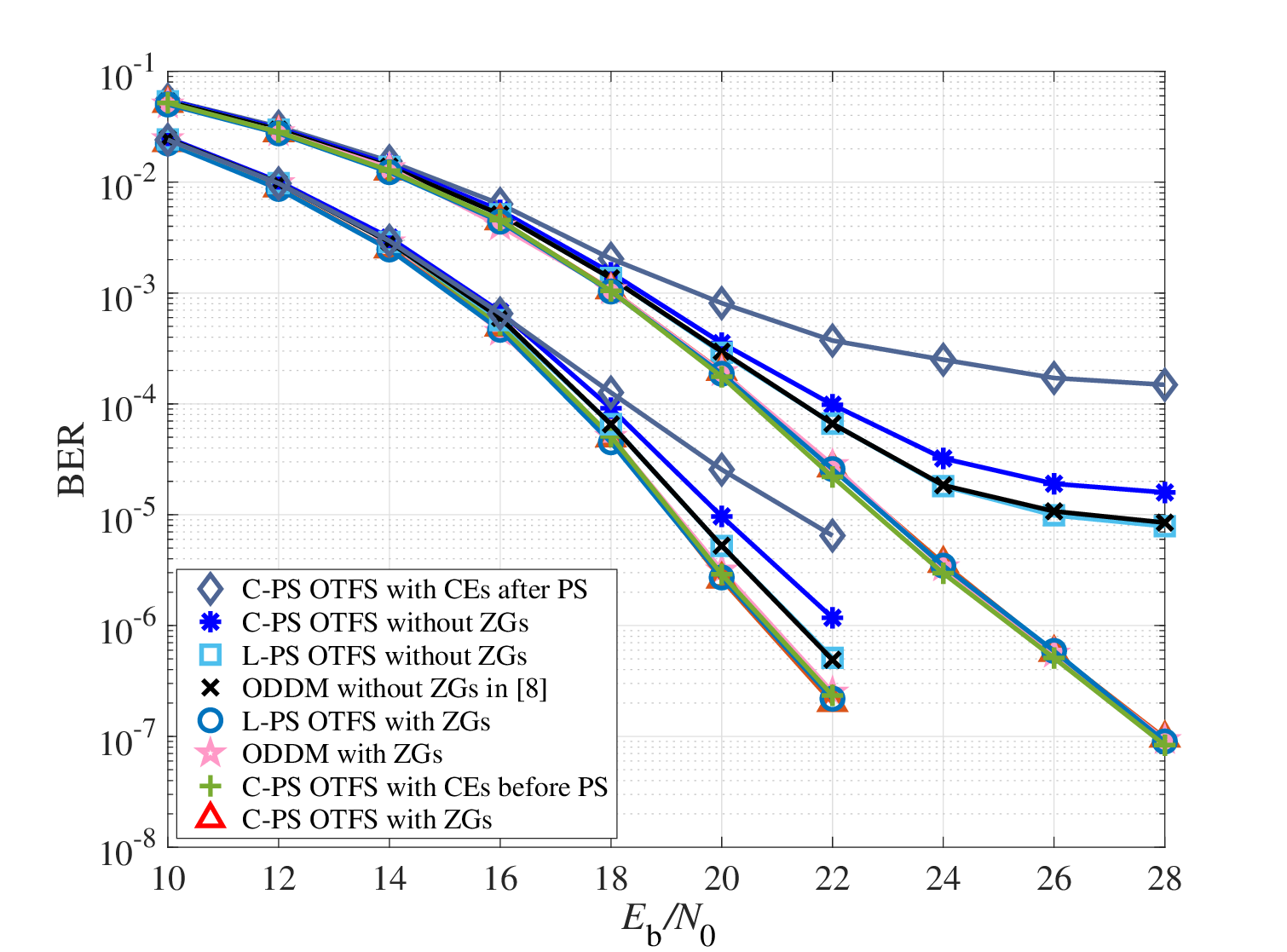}}
  \caption{BER performance of different pulse-shaping techniques for 4-QAM and 16-QAM at the relative velocity of $v=500$~km/h.}
  \label{fig:BER_v500}
    \vspace{-0.3cm}
\end{figure}

\begin{figure}[!t]
  \centering 
  {\includegraphics[scale=0.33]{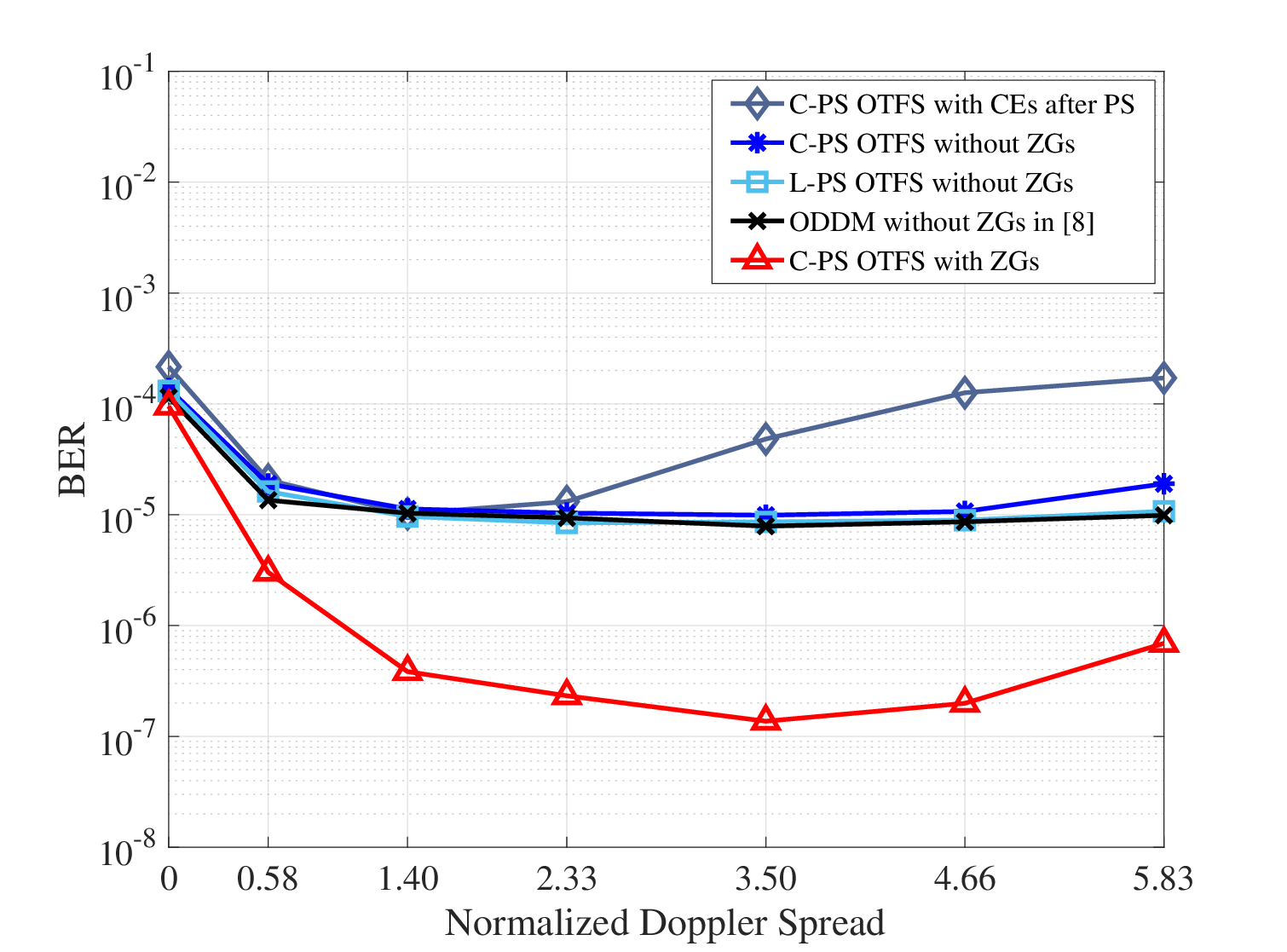}}
  \caption{Doppler spread effect on the BER performance of different pulse-shaping techniques at $E_{\rm{b}}/N_0=26~\rm{dB}$ for 16-QAM.}
  \label{fig:BER_v}
\end{figure}

\begin{figure}[!t]
  \centering 
  {\includegraphics[scale=0.33]{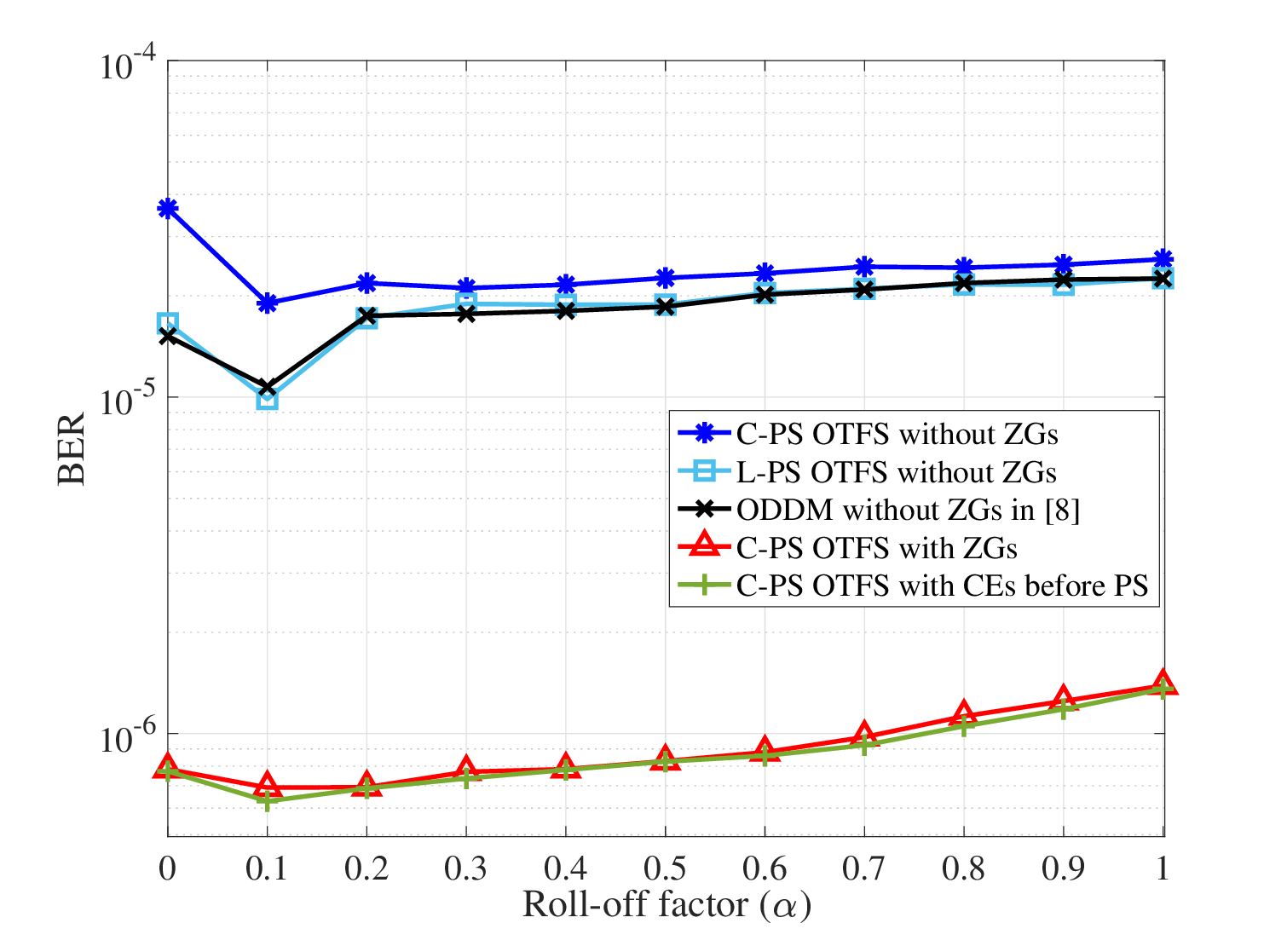}} 
    \vspace{-0.12cm}
  \caption{BER performance versus roll-off factor of the pulse-shaping filter for 16-QAM at $E_{\rm{b}}/N_0=26~\rm{dB}$ and the relative velocity of $v=500$~km/h.}
  \label{fig:BER_RF}
\end{figure}

The BER performance versus $E_{\rm{b}}/N_0$ for the pulse-shaping techniques under study, in two cases of 4-QAM and 16-QAM, is investigated in Fig.~\ref{fig:BER_v500}. As it was explained in Section~\ref{subsec:OOB reduction} and illustrated in Fig.~\ref{fig:BER_v}, CE addition and removal after pulse-shaping (PS) lead to the BER performance penalty in time-varying channels which escalates as the Doppler spread increases. 
In contrast, if the CEs are inserted along the delay dimension before oversampling and pulse-shaping, this issue is resolved and a significant BER performance improvement is achieved. Based on the results in Fig.~\ref{fig:BER_v500}, linear pulse-shaping techniques have a slightly better BER performance than circular pulse-shaping without windowing. However, inserting only a small number of ZGs along the delay dimension, i.e., $L_{\rm ZG}=6$, improves the performance of all the pulse-shaping techniques and brings them on par with each other. This improvement is more noticeable in case of 16-QAM, i.e., around 5~dB. As all the pulse-shaping techniques with ZGs have the same performance, in Figs.~\ref{fig:BER_v} and \ref{fig:BER_RF}, we only show the BER for C-PS OTFS with ZGs as an indicative result for  C/L-PS OTFS with ZGs. It is worth noting that inserting the CEs before PS for C-PS OTFS, has a similar effect to ZGs. 


To analyze the sensitivity of different pulse-shaping techniques to the channel variations, Fig.~\ref{fig:BER_v} evaluates the BER performance versus Doppler spread. For the sake of generality, Doppler spread is normalized to the Doppler spacing as $\nu_{\rm{max}}MNT_{\rm{s}}$. 
As shown, the performance for the linear and circular pulse-shaping techniques is improved up to a certain Doppler spread. This is due to the additional diversity gain that is introduced by the time-selectivity of the channel. However, further increase in Doppler spread leads to performance degradation.
For C-PS OTFS, CE insertion before or after pulse-shaping leads to about the same performance in low-mobility scenarios. However, based on the discussions in Section~\ref{subsec:OOB reduction}, as the Doppler spread increases, CEs must be inserted before pulse-shaping to avoid severe performance penalty.

Fig.~\ref{fig:BER_RF} illustrates the effect of various roll-off factors on circular and linear pulse-shaping techniques.
Increasing $\alpha$ results in the spread of the pulse-shaping filter in the frequency domain. This spreading effect leads to a slight degradation in the BER performance,
and the best performance is achieved when roll-off factor is around $\alpha=0.1$.

\begin{figure}[!t]
  \centering 
  {\includegraphics[scale=0.33]{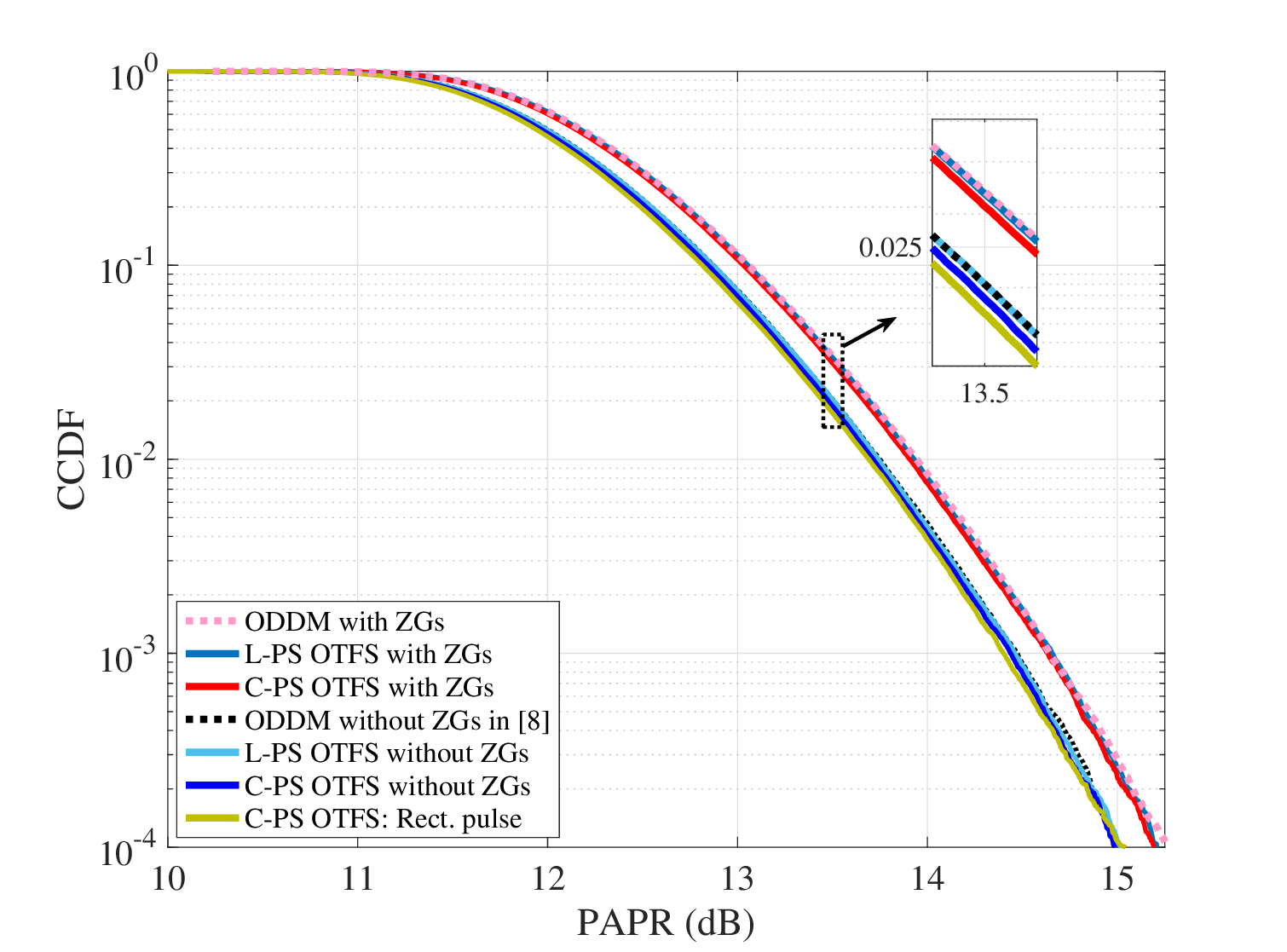}}
  \caption{PAPR performance comparison of different pulse-shaping techniques for 16-QAM.}
  \label{fig:CCDF}
\end{figure}
\begin{figure}[!t]
  \centering 
  {\includegraphics[scale=0.33]{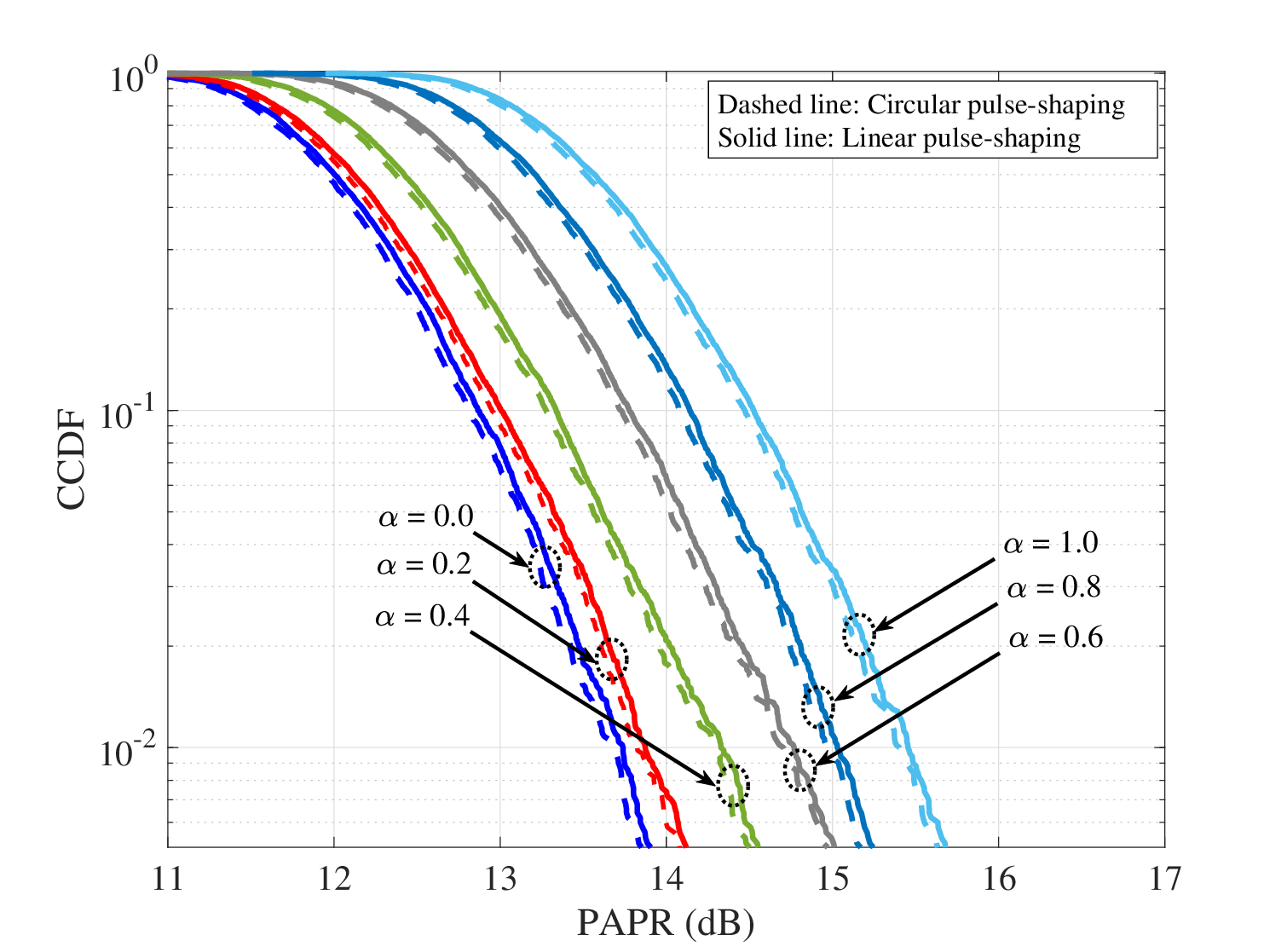}}
  \caption{Effect of Roll-off factor on PAPR performance of different pulse-shaping techniques for 16-QAM.}
  \label{fig:CCDF_RF}
\end{figure}

Fig.~\ref{fig:CCDF} represents the complementary cumulative distribution function (CCDF) of the PAPR for different pulse-shaping techniques on the delay-Doppler plane. These results demonstrate about the same PAPR performance for both the linear and circular pulse-shaping techniques. An observation here is that the ZG inclusion has the effect of a slight increase in PAPR as the ZGs lead to a reduced average signal power. According to Fig.~\ref{fig:CCDF}, ODDM and L-PS OTFS have the same PAPR. Thus, in Fig.~\ref{fig:CCDF_RF}, we only examine the influence of the roll-off factor on the PAPR of C-PS OTFS and L-PS OTFS in absence of ZGs. As shown, an increase in the roll-off factor corresponds to a higher PAPR in both C-PS and L-PS OTFS which is due to the reduced sidelobes of the pulse.

\section{conclusion}\label{sec:Conclusion}

In this paper, we categorized delay-Doppler plane pulse-shaping techniques into two classes of circular and linear pulse-shaping. 
This led to the development of a unified framework that encompasses different pulse-shaping techniques for delay-Doppler multiplexing. 
We also established the discrete-time formulation of the recently emerged waveform ODDM which revealed that it is a linear pulse-shaping technique. 
An interesting finding in this study was that ODDM deploys the modulated prototype filter for data transmission in each Doppler subcarrier which mathematically explained the reason behind its staircase spectral behaviour. Using our unified pulse-shaping framework, we derived a generalized input-output relationship that captures the impact of pulse-shaping in the equivalent channel. 
For the ease of derivations, in the existing literature, \cite{Raviteja2018b,Yao2021,Tong2023,Yogesh2023}, the transmit and receive pulse-shaping filters are merged which is only valid when the channel is time-invariant. Hence, for the accuracy of derivations, we avoid merging these filters. We also introduced a unified modem structure that led to the proposal of fast convolution based low-complexity modem structures. According to our complexity analysis, the proposed low-complexity implementation techniques in this paper lead to significant complexity reduction compared to the existing structures in literature. 
We also proposed effective techniques that reduce the OOB emissions of circularly pulse-shaped signals by around 20~dB.
Our proposed OOB emission reduction techniques require a small number of guard symbols at the edges of each delay block which also lead to BER performance improvement for both linear and circular pulse-shaping techniques. 
Using guard symbols, all the pulse-shaping techniques under study achieve the same BER performance. 
We also compared different pulse-shaping techniques using various performance metrics. Based on our results, C-PS OTFS with guard symbols seems to be a good candidate due to its low computational cost, reasonably low OOB emissions and similar BER and PAPR performance to other pulse-shaping techniques.

\bibliographystyle{IEEEtran} 
\bibliography{IEEEabrv,references}

\begin{thebibliography}{10}
\providecommand{\url}[1]{#1}
\csname url@samestyle\endcsname
\providecommand{\newblock}{\relax}
\providecommand{\bibinfo}[2]{#2}
\providecommand{\BIBentrySTDinterwordspacing}{\spaceskip=0pt\relax}
\providecommand{\BIBentryALTinterwordstretchfactor}{4}
\providecommand{\BIBentryALTinterwordspacing}{\spaceskip=\fontdimen2\font plus
\BIBentryALTinterwordstretchfactor\fontdimen3\font minus
  \fontdimen4\font\relax}
\providecommand{\BIBforeignlanguage}[2]{{%
\expandafter\ifx\csname l@#1\endcsname\relax
\typeout{** WARNING: IEEEtran.bst: No hyphenation pattern has been}%
\typeout{** loaded for the language `#1'. Using the pattern for}%
\typeout{** the default language instead.}%
\else
\language=\csname l@#1\endcsname
\fi
#2}}
\providecommand{\BIBdecl}{\relax}
\BIBdecl

\bibitem{Hadani2017}
R.~Hadani, S.~Rakib, M.~Tsatsanis, A.~Monk, A.~J. Goldsmith, A.~F. Molisch, and
  R.~Calderbank, ``Orthogonal time frequency space modulation,'' in \emph{IEEE
  Wireless Commun. and Netw. Conf.}, 2017, pp. 1--6.

\bibitem{Hadani2018}
\BIBentryALTinterwordspacing
R.~Hadani, S.~Rakib, S.~Kons, M.~Tsatsanis, A.~Monk, C.~Ibars, J.~Delfeld,
  Y.~Hebron, A.~J. Goldsmith, A.~F. Molisch, and R.~Calderbank, ``Orthogonal
  time frequency space modulation,'' 2018. [Online]. Available:
  \url{https://doi.org/10.48550/arXiv.1808.00519}
\BIBentrySTDinterwordspacing

\bibitem{Wei2021}
Z.~Wei, W.~Yuan, S.~Li, J.~Yuan, G.~Bharatula, R.~Hadani, and L.~Hanzo,
  ``Orthogonal time-frequency space modulation: A promising next-generation
  waveform,'' \emph{IEEE Wireless Commun.}, vol.~28, no.~4, pp. 136--144, 2021.

\bibitem{Tiwari2020}
S.~Tiwari and S.~S. Das, ``Circularly pulse-shaped orthogonal time frequency
  space modulation,'' \emph{IEEE Electron. Lett.}, vol.~56, no.~3, pp.
  157--160, 2020.

\bibitem{Pishva2022}
S.~M. Pishvaei, B.~M. Tazehkand, and J.~Pourrostam, ``Design and performance
  evaluation of {FBMC}-based orthogonal time–frequency space ({OTFS})
  system,'' \emph{Physical Commun.}, vol.~53, p. 101723, 2022.

\bibitem{Andrea2023}
\BIBentryALTinterwordspacing
C.~D'Andrea, S.~Buzzi, M.~Fresia, and X.~Wu, ``Doppler-resilient universal
  filtered multicarrier ({DR-UFMC}): A beyond-{OTFS} modulation,'' 2023.
  [Online]. Available: \url{https://arxiv.org/abs/2302.09405}
\BIBentrySTDinterwordspacing

\bibitem{Lin2022_1}
H.~Lin and J.~Yuan, ``Multicarrier modulation on delay-{D}oppler plane:
  Achieving orthogonality with fine resolutions,'' in \emph{IEEE Int. Conf. on
  Commun.}, 2022, pp. 2417--2422.

\bibitem{Lin2022_2}
------, ``Orthogonal delay-{D}oppler division multiplexing modulation,''
  \emph{IEEE Trans. on Wireless Commun.}, vol.~21, no.~12, pp.
  11\,024--11\,037, 2022.

\bibitem{Lin2023_2}
\BIBentryALTinterwordspacing
H.~Lin, J.~Yuan, W.~Yu, J.~Wu, and L.~Hanzo, ``Multi-carrier modulation: An
  evolution from time-frequency domain to delay-{D}oppler domain,'' 2023.
  [Online]. Available: \url{https://doi.org/10.48550/arXiv.2308.01802}
\BIBentrySTDinterwordspacing

\bibitem{Fred2018}
F.~Wiffen, L.~Sayer, M.~Z. Bocus, A.~Doufexi, and A.~Nix, ``Comparison of
  {OTFS} and {OFDM} in ray launched sub-6 {GH}z and mmwave line-of-sight
  mobility channels,'' in \emph{IEEE Annu. Int. Symp. on Pers., Indoor and
  Mobile Radio Commun.}, 2018, pp. 73--79.

\bibitem{Raviteja2019b}
P.~Raviteja, Y.~Hong, E.~Viterbo, and E.~Biglieri, ``Practical pulse-shaping
  waveforms for reduced-cyclic-prefix {OTFS},'' \emph{IEEE Trans. on Veh.
  Technol.}, vol.~68, no.~1, pp. 957--961, 2019.

\bibitem{Raviteja2018b}
P.~Raviteja, K.~T. Phan, Y.~Hong, and E.~Viterbo, ``Interference cancellation
  and iterative detection for orthogonal time frequency space modulation,''
  \emph{IEEE Trans. on Wireless Commun.}, vol.~17, no.~10, pp. 6501--6515,
  2018.

\bibitem{Zhou2023}
A.~Zhou, Y.~Pan, J.~Wu, H.~Lin, and J.~Yuan, ``On the performance of practical
  pulse-shaped {OTFS} with analog receivers,'' in \emph{IEEE Int. Conf. on
  Commun. Workshops}, 2023, pp. 518--523.

\bibitem{Lin2022_0}
C.~Shen, J.~Yuan, and H.~Lin, ``Error performance of rectangular pulse-shaped
  {OTFS} with practical receivers,'' \emph{IEEE Wireless Commun. Lett.},
  vol.~11, no.~12, pp. 2690--2694, 2022.

\bibitem{Yao2021}
Y.~Ge, Q.~Deng, P.~C. Ching, and Z.~Ding, ``Receiver design for {OTFS} with a
  fractionally spaced sampling approach,'' \emph{IEEE Trans. on Wireless
  Commun.}, vol.~20, no.~7, pp. 4072--4086, 2021.

\bibitem{Tong2023}
J.~Tong, J.~Xi, J.~Yuan, and H.~Lin, ``On the input-output relation of {ODDM}
  modulation over general physical channels,'' in \emph{IEEE Int. Conf. on
  Commun. Workshops}, 2023, pp. 289--294.

\bibitem{Wei2021_2}
Z.~Wei, W.~Yuan, S.~Li, J.~Yuan, and D.~W.~K. Ng, ``Transmitter and receiver
  window designs for orthogonal time-frequency space modulation,'' \emph{IEEE
  Trans. on Commun.}, vol.~69, no.~4, pp. 2207--2223, 2021.

\bibitem{Tusha2023}
A.~Tusha and H.~Arslan, ``Low complex inter-doppler interference mitigation for
  {OTFS} systems via global receiver windowing,'' \emph{IEEE Trans. on Veh.
  Technol.}, vol.~72, no.~6, pp. 7685--7698, 2023.

\bibitem{Farhang2018}
A.~Farhang, A.~RezazadehReyhani, L.~E. Doyle, and B.~Farhang-Boroujeny, ``Low
  complexity modem structure for {OFDM}-based orthogonal time frequency space
  modulation,'' \emph{IEEE Wireless Commun. Lett.}, vol.~7, no.~3, pp.
  344--347, 2018.

\bibitem{Franz2022}
F.~Lampel, A.~Avarad, and F.~M. Willems, ``On {OTFS} using the discrete {Z}ak
  transform,'' in \emph{IEEE Int. Conf. on Commun. Workshops}, 2022, pp.
  729--734.

\bibitem{Saifkhan2021}
S.~K. Mohammed, ``Derivation of {OTFS} modulation from first principles,''
  \emph{IEEE Trans. on Veh. Technol.}, vol.~70, no.~8, pp. 7619--7636, 2021.

\bibitem{Saifkhan2022}
S.~K. Mohammed, R.~Hadani, A.~Chockalingam, and R.~Calderbank, ``{OTFS}—a
  mathematical foundation for communication and radar sensing in the
  delay-{D}oppler domain,'' \emph{IEEE BITS the Inf. Theory Mag.}, vol.~2,
  no.~2, pp. 36--55, 2022.

\bibitem{Lin2022_3}
H.~Lin and J.~Yuan, ``On delay-{D}oppler plane orthogonal pulse,'' in
  \emph{IEEE Global Commun. Conf.}, 2022, pp. 5589--5594.

\bibitem{Li2023}
\BIBentryALTinterwordspacing
S.~Li, W.~Yuan, Z.~Wei, J.~Yuan, B.~Bai, and G.~Caire, ``On the pulse shaping
  for delay-{D}oppler communications,'' 2023. [Online]. Available:
  \url{https://doi.org/10.48550/arXiv.2306.08704}
\BIBentrySTDinterwordspacing

\bibitem{Wei2023}
W.~Yuan, S.~Li, Z.~Wei, Y.~Cui, J.~Jiang, H.~Zhang, and P.~Fan, ``New delay
  {D}oppler communication paradigm in 6{G} era: A survey of orthogonal time
  frequency space ({OTFS}),'' \emph{China Commun.}, vol.~20, no.~6, pp. 1--25,
  2023.

\bibitem{PS2007}
K.~Gentile, ``Digital pulse-shaping filter basics,'' Analoge Devices, AN-922,
  Application Note, 2007.

\bibitem{Farhang2010}
B.~Farhang-Boroujeny, \emph{Signal Processing Technique for Software
  Radios}.\hskip 1em plus 0.5em minus 0.4em\relax Lulu Publishing House, 2010.

\bibitem{proakis2007digital}
J.~Proakis and D.~Manolakis, \emph{Digital Signal Processing}.\hskip 1em plus
  0.5em minus 0.4em\relax Pearson Prentice Hall, 2007.

\bibitem{AFarhang2010}
A.~Farhang, M.~Molavi~Kakhki, and B.~Farhang-Boroujeny, ``Wavelet-{OFDM} versus
  filtered-{OFDM} in power line communication systems,'' in \emph{Int. Symp. on
  Telecommun.}, 2010, pp. 691--694.

\bibitem{Zegrar2023}
S.~E. Zegrar and H.~Arslan, ``A novel cyclic prefix configuration for enhanced
  reliability and spectral efficiency in {OTFS} systems,'' \emph{IEEE Wireless
  Commun. Lett.}, vol.~12, no.~5, pp. 888--892, 2023.

\bibitem{FarhangB2016}
B.~Farhang-Boroujeny, A.~Farhang, A.~RezazadehReyhani, A.~Aminjavaheri, and
  D.~Qu, ``A comparison of linear {FBMC} and circularly shaped waveforms,'' in
  \emph{IEEE/ACES Int. Conf. on Wireless Inf. Technol. and Syst. and Appl.
  Comput. Electromagn.}, 2016, pp. 1--2.

\bibitem{3gpp}
3GPP, ``Evolved universal terrestrial radio access ({E-UTRA}); base station
  ({BS}) radio transmission and reception,'' 3rd Generation Partnership Project
  (3GPP), TS 36.104 V15.3.0, 2018.

\bibitem{Yogesh2023}
V.~Yogesh, V.~S. Bhat, S.~R. Mattu, and A.~Chockalingam, ``On the bit error
  performance of {OTFS} modulation using discrete {Z}ak transform,'' in
  \emph{IEEE Int. Conf. on Commun.}, 2023, pp. 741--746.

\end{thebibliography}

\end{document}